\documentclass[preprint2]{aastex6}

\usepackage{amssymb,amsmath,graphicx,natbib,hyperref}
\usepackage{color}
\usepackage{enumerate}

\begin{document}

\title{SDSS-IV MaNGA: What shapes the distribution of metals in galaxies? Exploring the roles of the local gas~fraction and escape velocity}

\email{jbarrer3@jhu.edu}

\author{J.K. Barrera-Ballesteros\altaffilmark{1}, T. Heckman\altaffilmark{1}, S.F. S\'anchez\altaffilmark{2}, N.L. Zakamska\altaffilmark{1}, J. Cleary\altaffilmark{1}, G. Zhu\altaffilmark{1}, J. Brinkmann\altaffilmark{3}, N. Drory\altaffilmark{4} \& the MaNGA team}
\altaffiliation{Department of Physics \& Astronomy, Johns Hopkins University, Bloomberg Center, 3400 N. Charles St., Baltimore, MD 21218, USA}
\altaffiliation{Instituto de Astronom\'ia, Universidad Nacional Aut\'onoma de M\'exico, A.P. 70-264, 04510 M\'exico, D.F., M\'exico}
\altaffiliation{Apache Point Observatory, Sunspot, NM, 88349, USA}
\altaffiliation{McDonald Observatory, The University of Texas at Austin, 1 University Station, Austin, TX 78712, USA}

\begin{abstract}
We determine the local metallicity of the ionized gas for more than \mbox{9.2$\times$10$^{5}$} star forming regions (spaxels) located in 1023 nearby galaxies included in the SDSS-IV MaNGA IFU survey. We use the dust extinction derived from the Balmer decrement and stellar template fitting in each spaxel to estimate the local gas and stellar mass densities, respectively. We also use the measured rotation curves to determine the local escape velocity ($V_\mathrm{esc}$). We have then analyze the relationships between the local metallicity and both the local gas fraction ($\mu$) and $V_\mathrm{esc}$. We find that metallicity decreases with both increasing $\mu$ and decreasing $V_\mathrm{esc}$. By examining the residuals in these relations we show that the gas fraction plays a more primary role in the local chemical enrichment than $V_\mathrm{esc}$.  We show that the gas-regulator model of chemical evolution provides a reasonable explanation of the metallicity on local scales. The best-fit parameters for this model are consistent with metal loss caused by momentum-driven galactic outflows. We also argue that both the gas fraction and local escape velocity are connected to the local stellar surface density, which in turn is a tracer of the epoch at which the dominant local stellar population formed
\end{abstract}

\section{Introduction}

The observed gas-phase metal content in the interstellar medium (ISM) is the byproduct of stellar evolution, modulated by gas flows into and out of galaxies. Therefore, understanding how the metals relate to other observables provides key clues about how galaxies evolve and (more importantly) about the physical processes responsible for this evolution. 

In simple models of chemical evolution, the metal abundance in the ISM will set by three main processes. The first is the degree of chemical evolution, which increases as more gas is converted into massive stars, which return metal-enriched gas. The second is the loss of metals through outflows driven by the feedback effects of massive stars and supernovae. The fractional metal loss is expected to be larger in low mass galaxies with low escape velocities. The third is the inflow of lower metallicity gas from the surrounding circum-galactic medium. Indeed, recent analytical models suggest that chemical evolution can be understood as an interplay between these gas flows and the gas reservoir available to form new stars \citep{Lilly_2013}.

Over the last three decades many studies have been devoted to testing these models of galactic chemical evolution. One fundamental observational parameter in probing the chemical evolution of galaxies is their total gas fraction (i.e., the ratio between the gas and stellar plus gas masses). Gas-rich galaxies tend to show low metallicities in comparison to gas-poor massive ones \citep[e.g.,][]{Garnett_2002,Pilyugin_2004, 2013A&A...550A.115H, 2013MNRAS.433.1425B, 2016A&A...595A..48B}. This is qualitatively consistent with theoretical expectations. 

Different authors have also noted the correlation between a galaxy's luminosity and its oxygen abundances in samples of late type and irregular galaxies \citep[e.g.,][]{1992MNRAS.259..121V,1994ApJ...420...87Z,1979A&A....80..155L,Pilyugin_2004}. Using reliable estimates of the stellar mass and the very large data set provided by the Sloan Digital Sky Survey (SDSS), \cite{Tremonti_2004} showed a very tight relation between the stellar mass and the central gas-phase metallicity for more than 53,000 star-forming galaxies. The central metallicity increases with stellar mass, reaching a constant value for massive galaxies. This supports the idea of a mass-dependent loss of metals.

Due to the technical challenge in obtaining spatially resolved data for a large sample of galaxies, most of the studies to date have relied on integrated or central properties. In particular, the emission line fluxes used to derive the metallicity are usually obtained from the central region of the target which in turn only probe the physical processes that affect the metallicity in that specific region of galaxies. However, the situation has been changing significantly in the last years with the deployment of integral-field spectroscopy surveys. Thanks to them, different studies have unveiled relations at local scales that are analogous to those observed globally/centrally. \cite{Rosales-Ortega_2012} found a tight relation between the local metallicity and the local stellar surface mass density. This relation has also been observed in large surveys such as the CALIFA survey \citep{2014A&A...563A..49S}. For a large sample of disk MaNGA galaxies (650), we were able to reproduce both the observed radial metallicity gradients and the global mass-metallicity relation using a single relationship between the local values of the metallicity and stellar surface mass density \citep{BB_2016}. This result indicates that local gas-phase metallicity is a consequence of the local star formation history.

As noted above, two important quantities in models of chemical evolution are the gas fraction and the escape velocity. Studies exploring the spatially resolved metallicity as a function of the gas fraction and/or escape velocity are rather scarce. With a large, homogeneous dataset from the GASS survey \cite{2012ApJ...745...66M} found that the metallicity in the outskirts of a sample of star forming galaxies decreases as their total atomic gas fraction increases, suggesting gas fraction is fundamentally important. More recently, \cite{Carton_2015} studied the metallicity gradients in a sample of 50 late-type galaxies. In order to explain the observed radial distributions, the authors used the gas fraction in a local version of a chemical model in which inflows, outflows and star formation are in equilibrium \citep{Lilly_2013}. The gradients are then explained by the radial variation of the gas fraction and the mass loading factor (the ratio of the outflow and the star formation rate), which is in turn driven by the radial variation in the escape velocity. On the other hand, \cite{Ho_2015} explained the similarity in the observed metallicity gradients of 49 late-type galaxies by assuming a model in which the mass loading factor and the ratio of the inflows and star formation rates are constant. Their results suggest that at local scales the chemical evolution of galaxies is virtually the same as a closed-box model in which the metallicity depends only on the local gas fraction, and very weakly on inflows or outflows. 

In this paper our goal is to exploit the data from the Sloan Digital Sky Survey in the field of spatially resolved spectroscopy via the MaNGA survey \citep{Bundy_2015} in order to study the impact of the local gas fraction and the local escape velocity on the local metallicity in more than 1700 star-forming galaxies. The motivation is to provide insights into the physical processes responsible for chemical evolution on local scales and to better understand the origin and meaning of the corresponding global relations. The structure of this paper is the following: in Section~\ref{sec:Sample} we introduce the main aspects of the MaNGA survey as well as the selection criteria for our sample; in Section~\ref{sec:Analysis} we derive the spatially resolved quantities that we used in the study while in Section~\ref{sec:Results} we present the relations between these parameters; in Section~\ref{sec:Models} we compare those relations with models of chemical evolution, and discuss the implications of our findings in Section~\ref{sec:Discu}; finally in Section~\ref{sec:Concl} we present our conclusions. 

\section{Sample and data}
\label{sec:Sample}
\subsection{The MaNGA sample and datacubes}
For this study we use the sample of galaxies observed by the MaNGA survey through June 2016. The goal of this ongoing survey is to observe a sample of 10,000 nearby galaxies using an integral field spectroscopy unit \citep[IFU,][]{Bundy_2015}. The MaNGA survey is taking place at the 2.5 meter Sloan Telescope at the Apache Point Observatory \citep{2006AJ....131.2332G}. Observations are carried out using a set of 17 different fiber-bundles (science IFUs) packed in hexagon shapes \citep{2015AJ....149...77D}. The number of fibers varies from 19 to 127 per bundle covering a Field-of-View (FoV) of 12 to 32 arcsec, respectively. The diameter of the fibers in each of these bundles is 2.7 arcsec. These bundles feed two dual channel spectrographs covering a large wavelength range from 3600\ to 10000 \AA\, and provide a spectral resolution of $R \sim$ 2000 \citep{2013AJ....146...32S}. Details of the spectrophotometric calibrations for the MaNGA survey can be found in \citep{2016AJ....151....8Y}. The observing strategy includes a three-point dithering in order to provide homogeneous coverage of the field of view.  Final datacubes are reduced by a dedicated pipeline described in \cite{2016AJ....152...83L}. The pipeline accounts for sky subtraction, wavelength and flux calibration, as well as the combination of the three dithered observations. The final product of this pipeline is a datacube where each element is described by two spatial coordinates $x$ and $y$ corresponding to the RA and DEC projected on the sky and the $z$ coordinate corresponding to the wavelength element. Each of the spatial elements containing individual spectra are also known as spaxels. The spatial size of each spaxel in the final datacube is 0.5 arcsec.

The targets for MaNGA observations have been selected from the extended NASA-Sloan catalogue \citep[NSA \footnote{http://www.nsatlas.org}][]{2011AJ....142...31B}. This catalogue provides a large set of spectro-photometric parameters for individual targets, such as redshift ($z$), total stellar mass ($\mathrm{M_{*}}$), half-light radius ($\mathrm{R_{eff}}$), absolute $ugriz$ magnitudes, photometric position angle (PA), photometric ellipticity ($\epsilon$), among others. The main selection criteria for the MaNGA set of galaxies yield nearby targets with stellar masses $\mathrm{M_{*}} > 10^{9}\mathrm{M_{\odot}}$, a relatively flat distribution in stellar mass, and uniform radial coverage. Detailed descriptions of the selection parameters can be found in \cite{Bundy_2015}. To accomplish different science goals, the radial coverage of the IFU for $\sim$ 66\% of the sample is at least 1.5 $\mathrm{R_{eff}}$ (also known as the 'primary' sample) whereas for $\sim$ 30\% the IFU FoV covers at least 2.5 $\mathrm{R_{eff}}$ ('secondary' sample). For a description of the sample properties see \cite{2016AAS...22733401W}. The sample from which our sub-sample is drawn includes 2780 galaxies at redshift 0.01$< z <$0.17, covering a wide range of galaxy parameters (e.g, stellar mass, colors and morphology). For internal distribution purposes, this sample is known as the MPL-5. Most of these galaxies are included in the SDSS-IV DR14 data release \footnote{http://www.sdss.org/dr14/manga/}. These objects provide a panoramic view of the properties of the galaxy population in the Local Universe.

\subsection{Selected Sample}
\label{sec:Sample}
%
\begin{figure}
\includegraphics[width=\linewidth]{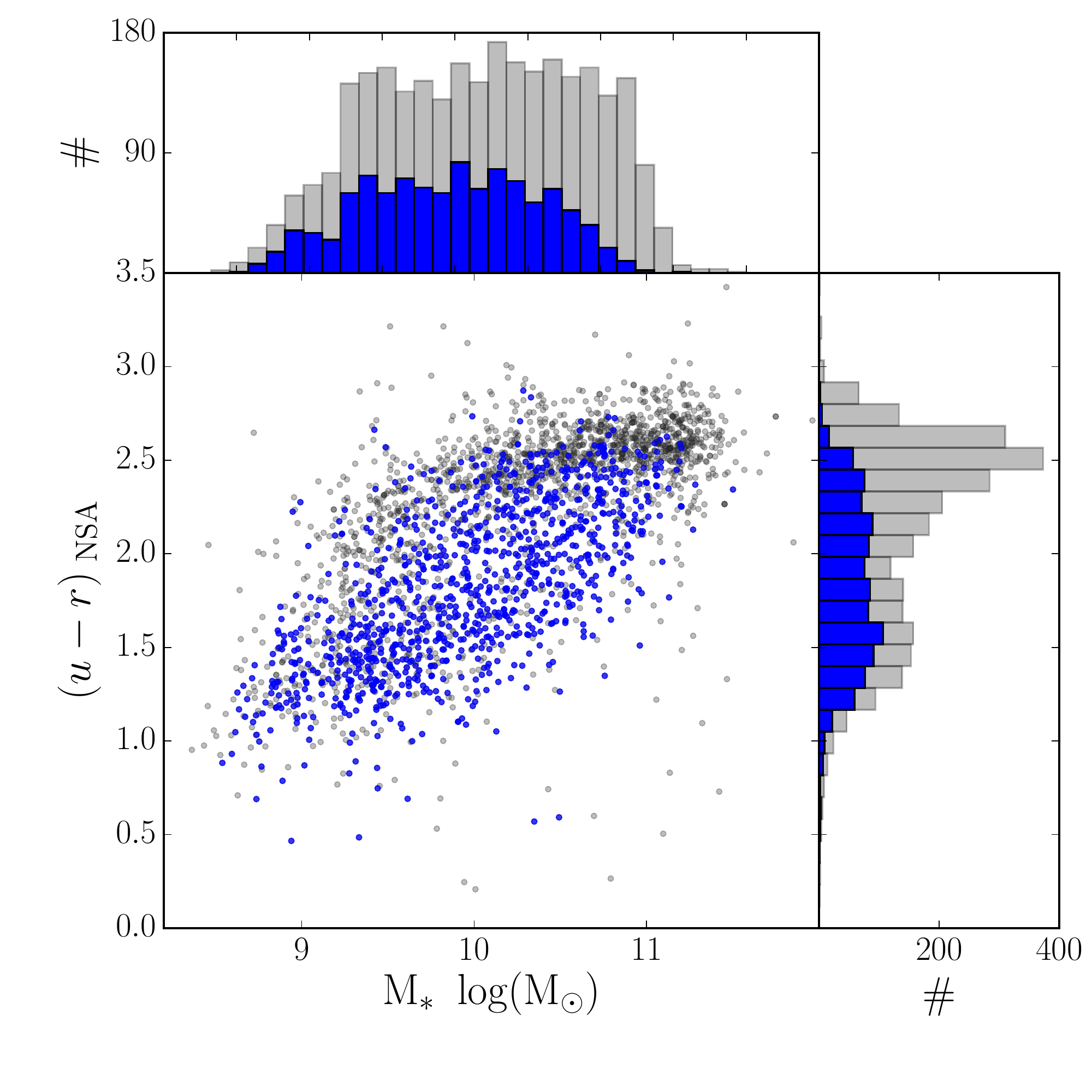}
\caption{Color-$\mathrm{M_{\ast}}$ diagram of the MPL-5 MaNGA sample. Gray and blue points represent the parent MPL-5 and our sample, respectively (for selection parameters see details in Secs.~\ref{sec:pipe3D} and ~\ref{sec:Vesc} ). The top-horizontal and right-vertical panels show the distributions for both samples (gray and blue histograms for the MPL-5 and selected samples, respectively) in stellar mass and color, respectively. The color distribution from our selected sample is flat in contrast to the observed bi-modality in the MPL-5 sample.} 
\label{fig:Sample}
\end{figure}

For this study we perform a further target selection using the following criteria: {\it (i)} galaxies with a representative number of spaxels ($>$ 10\%) classified as star forming (see Sec.~\ref{sec:pipe3D} for details) and {\it (ii)} reliable estimation of the rotation curve from their H$\alpha$ velocity field (see details in Sec.~\ref{sec:Vesc}). This selection yields a final sample that includes 1023 galaxies. In Fig.~\ref{fig:Sample} we compare the distribution of the selected sample against the entire MPL-5 sample in the color-$\mathrm{M_{\odot}}$ diagram.
The mass distribution of the MPL-5 sample is relatively flat in the mass range 9.2 $\lesssim\,\log(\mathrm{M_{*}/M_{\odot}}\,)\lesssim$ 11.0. The resulting mass distribution from our selection parameters mimics the MPL-5 parent flat distribution for a large range of stellar masses. Note however, that our sample does not cover the most massive galaxies bins. This is expected since from our selection parameters we required galaxies to have a significant amount of star-forming regions, which are not likely to occur in the most massive (early-type) galaxies.

It is evident that the MPL-5 sample follows the expected bimodality in color distributions (see gray diagram in the vertical panel in Fig.~\ref{fig:Sample}). However, our sample of galaxies shows a rather flat distribution over a wide range of colors, from blue to red. Our sample then includes not only star-forming galaxies (the so-called `blue cloud'), but also intermediate `green-valley' galaxies as well as `red-sequence' objects. The distribution of our sample in this diagram indicates that IFU spatially resolved data allows us to identify a significant fraction of galaxies that otherwise would not be classified as having star-forming regions.
\section{Analysis}
\label{sec:Analysis}

\subsection{Data Analysis Products from PIPE3D}
\label{sec:pipe3D}
\begin{figure*}
\includegraphics[width=\linewidth]{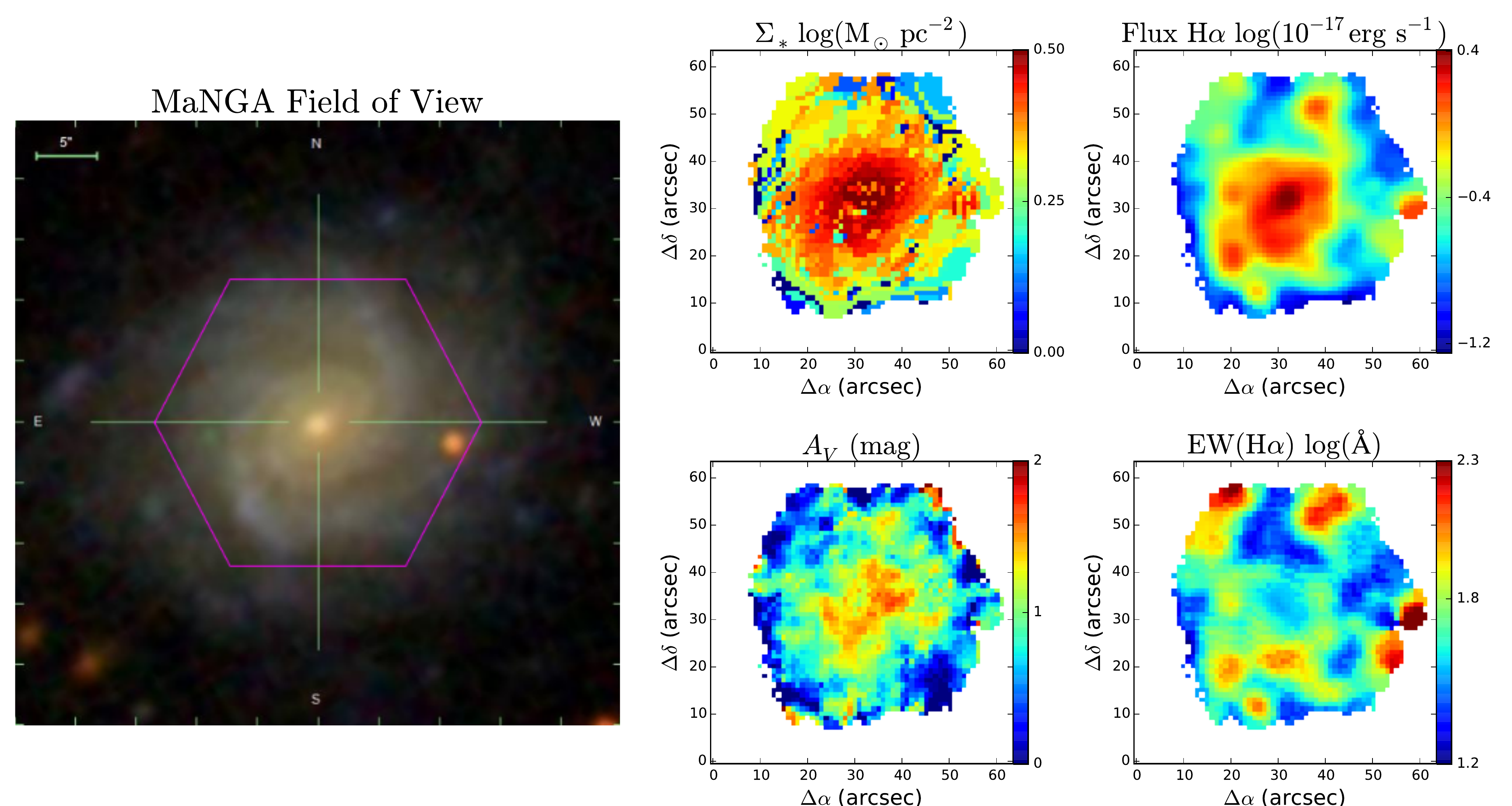}
\caption{Example of some of the maps of properties derived from the {\sc Pipe3D} analysis pipeline. The left image shows the hexagonal field of view covered by the MaNGA IFU over plotted in a composite SDSS image of one of the galaxies in our sample. In the right side of the figure, we show from left to right and top to bottom the maps for the surface mass density, the H$\alpha$ flux, the extinction, and the equivalent width of the H$\alpha$ emission line.} 
\label{fig:Example}
\end{figure*}
For our study we require the use of an analysis pipeline allowing us to extract spatially resolved maps in individual datacubes of the different properties of the ISM. For this we used the {\sc Pipe3D} analysis pipeline \citep{2016arXiv160201830S}. The pipeline provides a simultaneous fit of the reduced spectra using stellar population models as well as the nebular emission for the datacubes. To fit the stellar continuum the pipeline uses a single stellar population library from \cite{cid-fernandes13}. This library is a collection of 156 spectra that covers 39 stellar ages (from 1 to 3 Myrs), and four metallicities ($Z/Z_{\odot}$ = 0.2, 0.4, 1 and 1.5). Prior to the fitting of the stellar population, a spatial binning is performed to each data cube to reach a continuum S/N of 50 across the field of view. Then, the stellar population fitting is done for each of the coadded spectra. The stellar population for each spaxel within a bin is obtained by scaling the best fitted spectra derived for the bin to the continuum flux intensity of the corresponding spaxel. The pipeline also creates a continuum-free datacube by substracting the modeled continuum from the original datacube. The stellar mass map is determined by assuming the same extinction and mass to light ratio for all the spaxels within a given spatial bin. We determine the surface mass density ($\Sigma_*$) in each spaxel and correct from projection effects following the same procedure as in \cite{BB_2016}. 

The pipeline also fits single Gaussians to the strongest emission lines in the continuum-free datacube on a spaxel-by-spaxel basis. Thus, for a given emission line, the pipeline produces spatially resolved maps of the flux intensity, line-of-sight velocity, velocity dispersion, and equivalent width. For this study we use the derived flux maps from [OIII]$\lambda$5007, H$\beta$, H$\alpha$, and [NII]$\lambda$6548 as well as the velocity map from H$\alpha$. In Fig.~\ref{fig:Example} we show an example of the maps obtained using this analysis pipeline. 

We further select galaxies with star-forming spaxels. First we select spaxels with H$\alpha$ flux error smaller than 30 percent. Then, we use the classical BPT diagnostic diagram \citep{1981PASP...93....5B} that compares the line ratios [OIII]/ H$\beta$ and [NII]/ H$\alpha$. We classify as star forming those spaxels with line ratios below the Kauffmann demarcation line \citep{2003MNRAS.341...33K} and having EW(H$\alpha$) $>$ 6\AA. These two criteria provide a reliable selection of star forming regions allowing us to determine the metallicity indicators using HII-regions studies as well as the star formation rate. As result, our sample of star forming regions includes $\sim$ 9.2~$\times$~10$^{5}$ spaxels in the sample of 1023 galaxies. We use the abundance calibrator O3N2 from \citep{2004MNRAS.348L..59P}. This calibrator is based on the above line ratios
\begin{equation}
12+\log(\mathrm{O/H}) = 8.73 - 0.32 \times \mathrm{O3N2}
\end{equation}
with O3N2 = $\log$([OIII]/ H$\beta$) - $\log$([NII] / H$\alpha)$. The errors in the metallicity using this calibrator are typically of the order of $\sim$ 0.06 dex. 

\subsection{Gas Surface Mass Densities}
\label{sec:sig_gas}
\begin{figure*}
\includegraphics[width=\linewidth]{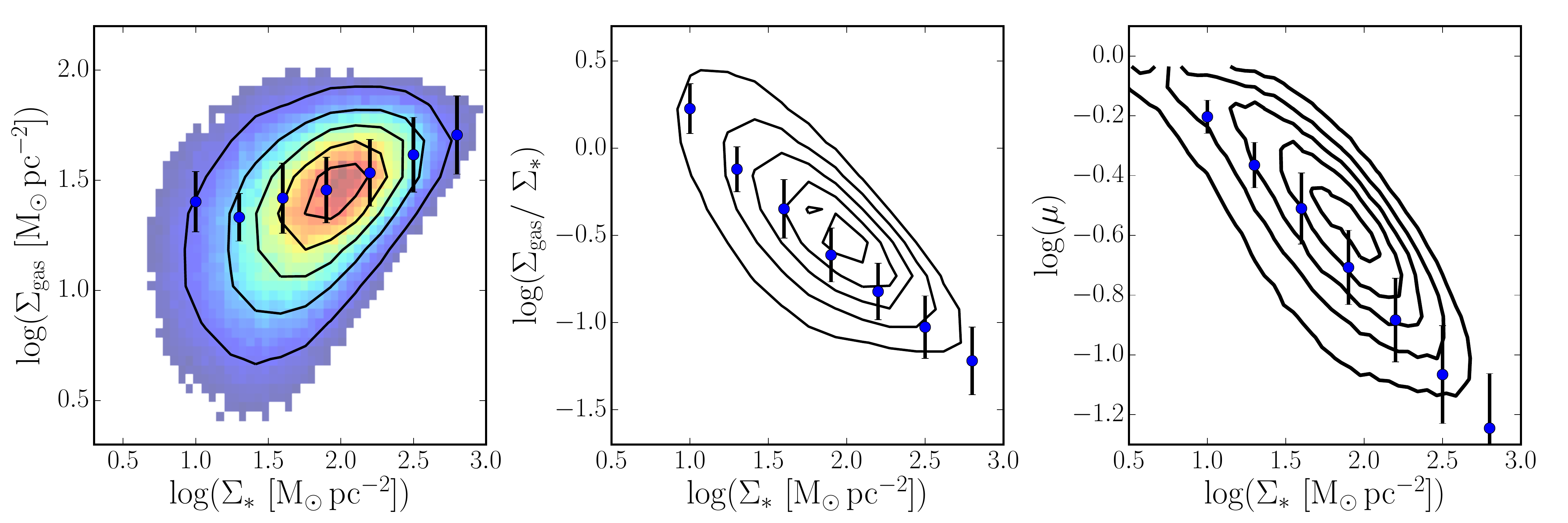}
\caption{Comparison of the estimation of  $\Sigma_\mathrm{gas}$ from the effective optical depth. \textit{Left panel:}  $\Sigma_\mathrm{gas}$ against  $\Sigma_\mathrm{*}$. \textit{Middle  panel:} Gas fraction ($\Sigma_\mathrm{gas}$/$\Sigma_\mathrm{*}$.) against  $\Sigma_\mathrm{*}$.  \textit{Right panel:} Total gas fraction ($\mu$) against  $\Sigma_\mathrm{*}$. For the three panels, the black solid contours represent the gas density in an effective screen framework. For each of these distributions,  the outer and innermost contours enclose 80\% and 10\% of the distribution, respectively.  In the left panel the distribution is also described by the color-coded distribution. The blue points and error bars represent the median and standard deviation of gas density $\Sigma_\mathrm{gas} = \Sigma_\mathrm{mol} +\Sigma_\mathrm{H_I}$ at different bins of $\Sigma_\mathrm{*}$. $\Sigma_\mathrm{mol}$ is derived from CO measurements in a sample of galaxies included in the EDGE-CALIFA survey \citep{Bolatto_2017}; we assume $\Sigma_\mathrm{H_I} = 10 {\rm M_{\odot} pc{-2}}$. These densities are determined for regions with CO flux larger than 2-$\sigma$ (Barrera-Ballesteros et al., in prep.). From these comparisons, the effective screen method seems to provide a similar estimation of the gas density as those provided from direct measurements of the molecular gas. } 
\label{fig:SigmaGasComp}
\end{figure*}

One of the main goals of this article is to understand the role of gas flows and of their relationship with the stellar component in shaping the gas-phase metallicity at local scales. We require an estimation of the total gas mass density ($\Sigma_\mathrm{gas}$). The most common observational technique to gauge the gas content in its different phases is at millimeter wavelengths. In particular, the emission line from the ground rotational transition  J = 1-0 of the CO molecule has been widely used as a tracer for H$_2$ molecular hydrogen in our Galaxy as well as in extragalactic sources \citep[for a recent  review see][]{Bolatto_2013}. However, direct spatially resolved observations of cold gas emission lines in large samples of galaxies, such as the MaNGA one, are technically quite challenging.   

From optical data, there are different methods or relations to estimate $\Sigma_\mathrm{gas}$. Among them, these methods include the inverse of the well-known Kennicutt-Schmidt Law \citep{Kennicutt_1998} which relates the star formation surface density ($\Sigma_\mathrm{SFR}$) to $\Sigma_\mathrm{gas}$; the relation with stellar mass density ($\Sigma_\mathrm{*}$) and  $\Sigma_\mathrm{gas}$; or assuming a constant ratio between gas column density and dust (we discuss these relations elsewhere, Barrrera-Ballesteros in prep.). In this study we simply adopt a full mixed distribution of gas and dust across an effective foreground screen that includes half of the gas column density \citep[e.g.,][]{McLeod_1993,Wuyts_2011,Imara_2007,Genzel_2013}. In this simple scenario, the total gas density is related to the optical extinction ($A_V$) via:
\begin{equation}
\Sigma_\mathrm{gas} = 30 M_{\odot}\,\,\mathrm{pc^{-2}} A_V
\label{eq:tau_tot}
\end{equation}

We obtain the optical extinction from the Balmer decrement. We derive the dust attenuation for the H$\alpha$ emission line [$A(H\alpha)$] following equation (1) from \cite{Catalan_Torrecilla_2015} using the H$\alpha$/H$\beta$ flux ratio, with its canonical Case B value of 2.86 and the extinction curve from \citep[][ $R_V$ = 3.1]{Cardelli_1989}.
\begin{equation}
A(H\alpha) = \frac{K_{H\alpha}}{-0.4(K_{H\alpha} -K_{H\beta} )} \times \log\left(\frac{F_{H\alpha}/F_{H\beta}}{2.86}\right)
\end{equation}
where$F_{\mathrm{H}\alpha}/F_{\mathrm{H}\beta}$ is the flux ratio between these Balmer lines, $K_{\mathrm{H}\alpha}$ = 2.53 and $K_{\mathrm{H}\beta}$ = 3.61 are the extinction coefficients for the Galactic extinction curve from \cite{Cardelli_1989}. \cite{Catalan_Torrecilla_2015} also noted that this attenuation is robust among different extinction curves and dust-to-star geometries (e.g. $R_V$ = 3.1 from \cite{Cardelli_1989} or $R_V$ = 4.05 from \cite{Calzetti_2000}). Using the above extinction curve, $A_V$ is simply $A_V$ = 0.83 $A(\mathrm{H}\alpha)$. 

We plot in Fig.\ref{fig:SigmaGasComp} the distribution of  $\Sigma_\mathrm{gas}$, the gas fraction ($\Sigma_\mathrm{gas} /  \Sigma_\mathrm{*}$) and the total gas fraction ($\mu~=~\Sigma_\mathrm{gas}~/~(\Sigma_\mathrm{gas}~+~\Sigma_\mathrm{*})$) against the $\Sigma_{*}$. In each panel, we over plot the median values of $\Sigma_{\rm gas} = \Sigma_{\rm mol}+\Sigma_{\rm H_I}$ for different $\Sigma_\mathrm{*} $ bins. $\Sigma_{\rm mol}$ comes from direct CO measurements from the EDGE survey \citep{Bolatto_2017}. We assume a constant distribution of $\Sigma_{\rm H_I} = 10 {\rm M_{\odot} pc^{-2}}$. In  Barrera-Ballesteros et al. (in prep.) we present a detailed study of the relations between molecular gas densities and spatially-resolved optical properties. This comparison indicates that the effective screen method provides a reliable estimation of   $\Sigma_\mathrm{gas} $ with similar values as those derived directly from CO measurements. 

\subsection{Rotation curve from Ha velocity fields}
\label{sec:HaVel}
\begin{figure*}
\includegraphics[width=\linewidth]{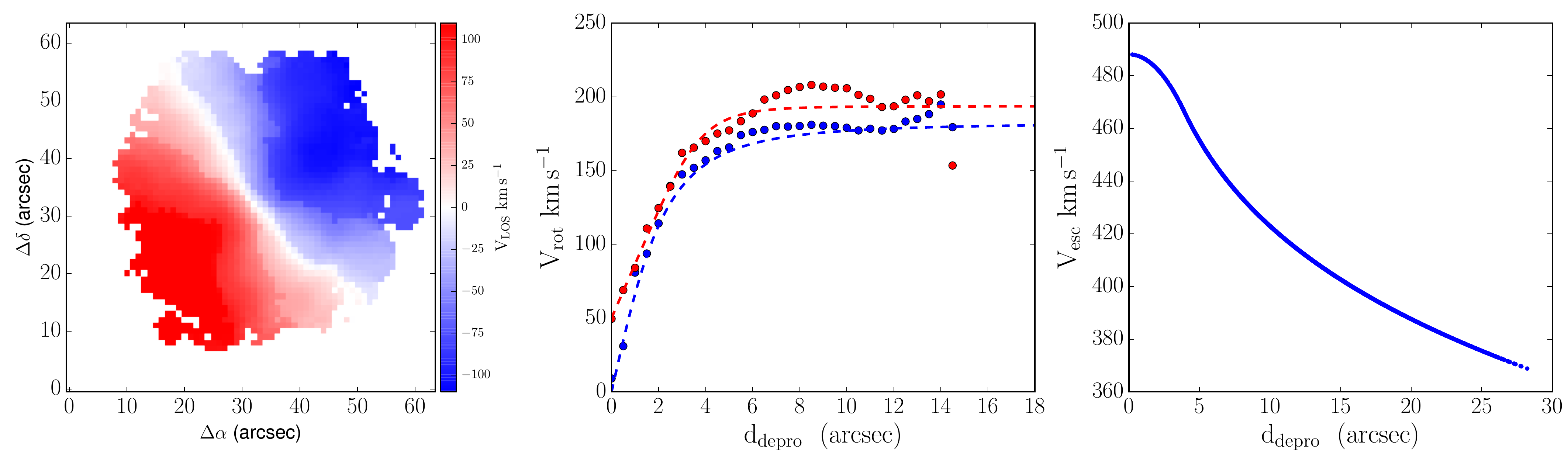}
\caption{Example of kinematic properties extracted from the MaNGA data cubes. The left panel shows the line of sight velocity map from the H$\alpha$ emission line. The middle panel shows the rotation curve derived from the velocity field for the approaching and receding sides (blue and red points, respectively; see details in Sec.~\ref{sec:HaVel}). The dashed lines represent the best fit of the Eq.~\ref{eq:v_rot} to the datapoints. The right panel shows the radial profile of the escape velocity $V_\mathrm{esc}$ (see details in Sec.~\ref{sec:Vesc}).} 
\label{fig:Example_kin}
\end{figure*}

\begin{figure}
\includegraphics[width=\linewidth]{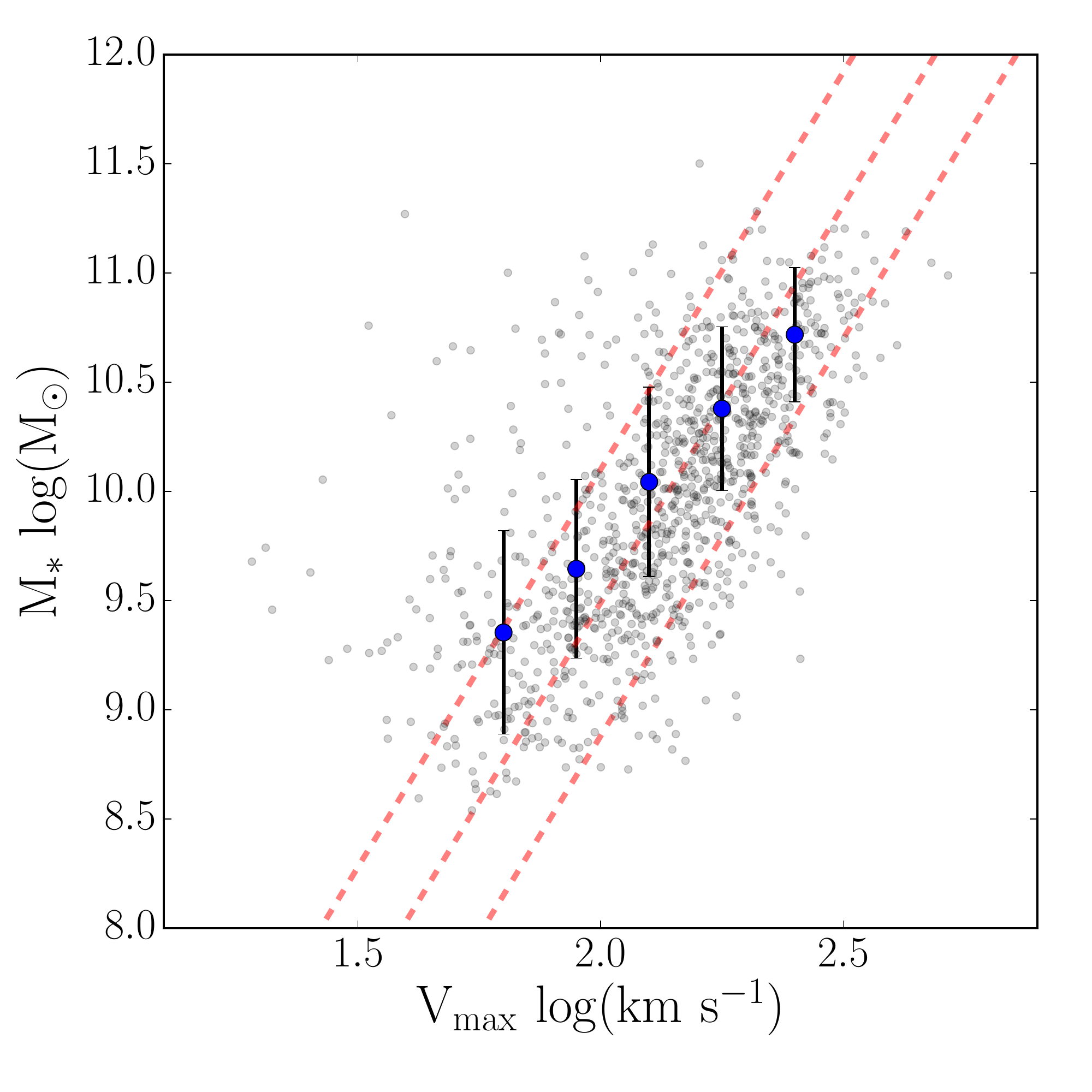}
\caption{Tully-Fisher relation for the selected sample of MaNGA galaxies. Blue points with error bars represent the medians and standard deviations stellar masses in bins of  50 $\mathrm{km\,\,s^{-1}}$ width. Red dashed lines represent the best values of the Tully-Fisher relation derived using H$\alpha$ velocity fields from the Fabry-Perot  GHASP survey \citep{TorresFlores_2011}. All the medians and most of the data points lie within the expected relation derived from this survey. } 
\label{fig:Sample_TF}
\end{figure}

As a first step in determining the local escape velocity for each of the star-forming spaxels in our sample of galaxies, we estimate the maximum rotation velocity ($V_\mathrm{max}$). To do so, we obtain the velocity curve for each of them via the velocity field from the H$\alpha$ emission line provided by the analysis pipeline (see an example in left panel of Fig.~\ref{fig:Example_kin} and details in Sec.\ref{sec:pipe3D}). We follow a similar procedure to that introduced in \cite{GarciaLorenzo_2015} and applied to a sample of CALIFA disk galaxies in \cite{BB_2014}. Using the H$\alpha$ velocity maps we determine the position of those spaxels with the maximum/minimum (receding/approaching) de-projected line-of-sight velocities compared to the systemic velocity at different de-projected distances (i.e., $V_\mathrm{rot}$ {\it vs} $d_\mathrm{depro}$, see red and blue dots in middle panel of Fig.~\ref{fig:Example_kin}). We assume that the kinematic center (i.e., position of the systemic velocity) is located at the optical nucleus. This is a reasonable assumption for disk galaxies where the location of the maximum gradient coincides in most of the cases with the optical nucleus \citep[e.g.,][]{BB_2014}. To de-project both the line-of-sight velocities and distances in our sample, we assume the photometric inclination from the NSA catalogue to be the kinematic inclination across the entire galaxy. Then, we parametrize our rotation curve as
\begin{equation}
V_\mathrm{rot} (d_\mathrm{depro}) = V_\mathrm{max} \cdot \frac{d_\mathrm{depro}}{(R^{\alpha}_\mathrm{turn} + d^{\alpha}_\mathrm{depro})^{1/\alpha}} 
\label{eq:v_rot}
\end{equation}
where $R_\mathrm{turn}$ is the galactocentric distance at which the rotation curve transitions from solid-body to flat (see Sec.~\ref{sec:Vesc}). We use this parametrization to fit the observed rotation curve to determine $V_\mathrm{max}$ (see dashed-lines in Fig.\ref{fig:Example_kin}).

As a sanity check, we investigate how well we are able to reproduce the observed Tully-Fisher relation using these derived velocities under the simple assumptions used here. In Fig.~\ref{fig:Sample_TF}, we plot $\log(V_\mathrm{max})$ against $\log(\mathrm{M_{\ast}/M_{\odot}}$ for those galaxies where we have reliable estimates of $V_\mathrm{max}$ (1023 galaxies) with uncertainties from the fitting smaller that 50 $\mathrm{km\,\,s^{-1}}$ (gray points). 
We also overplot in Fig.~\ref{fig:Sample_TF} with red-dashed lines the best-fit Tully-Fisher relation derived using Fabry-Perot H$\alpha$ velocity fields from a sample of field galaxies included in the GHASP survey \citep{TorresFlores_2011}. 
All the medians at different stellar mass bins and most of the individual values of $V_\mathrm{max}$ lie within the Tully-Fisher relation derived from the GHASP survey. At low stellar masses ($< 10^{9.5}$ M$_{\odot}$) we seem to slightly underestimate $V_\mathrm{max}$ (by $\sim 0.1$ dex). We also note that this is a mass range in which one must consider the velocity dispersion in the budget to account for the dynamical mass \citep[e.g.][]{Simons_2015}. Despite these limitations, the great majority of stellar masses and $V_\mathrm{max}$ in our sample are well described by the values from the best Tully-Fisher relation presented in \citep{TorresFlores_2011}.

\subsection{Escape velocity maps}
\label{sec:Vesc}
\begin{figure}
\includegraphics[width=\linewidth]{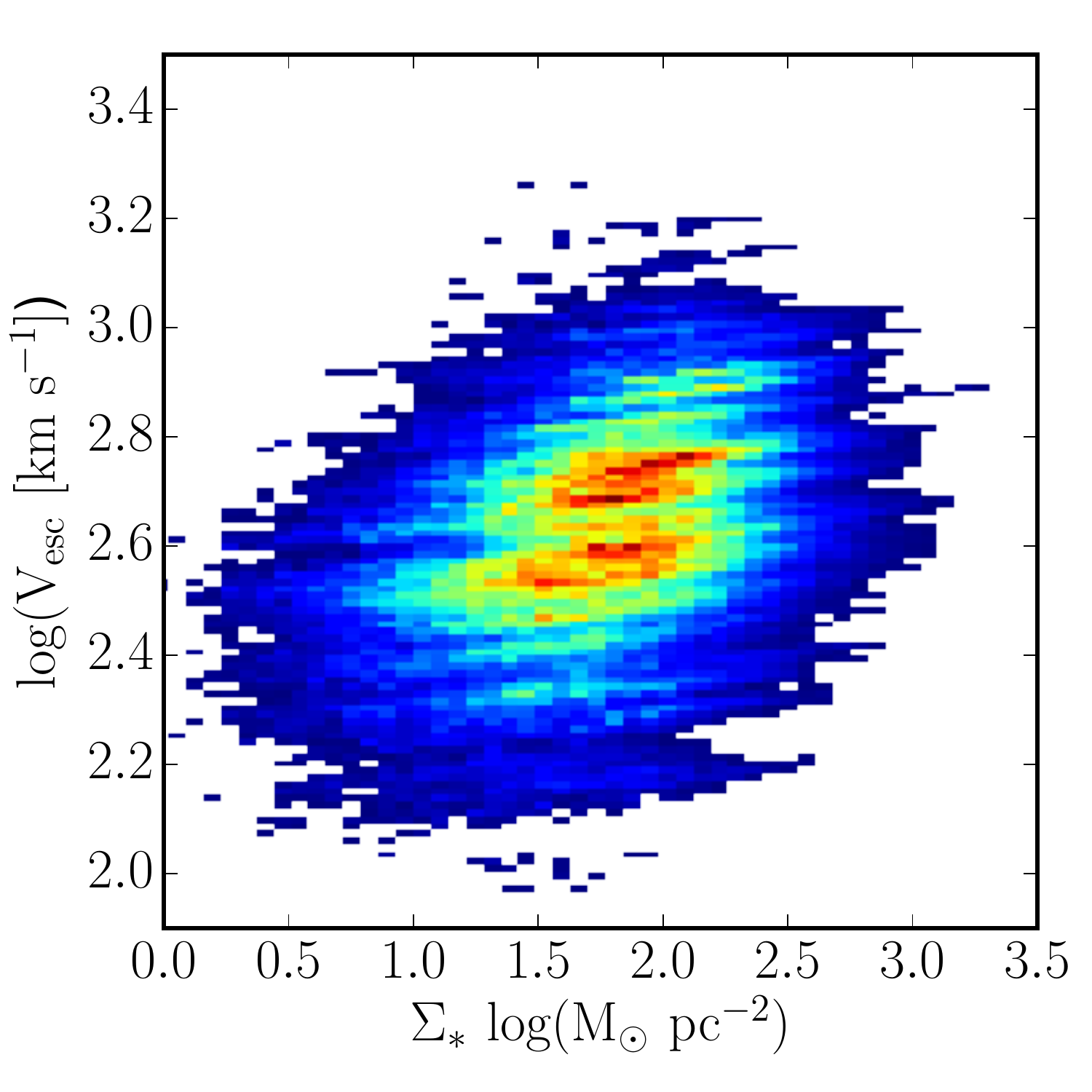}
\caption{Distribution of $V_\mathrm{esc}$ against $\Sigma_{*}$for our sample of star forming spaxels. Our sample spans a dynamic range of roughly an order-of-magnitude in $V_\mathrm{esc}$.} 
\label{fig:Vesc_Sigma}
\end{figure}

The amount of gas (and metals) expelled by the kinetic energy or momentum supplied by stellar feedback should depend on the depth of the local potential well. We have therefore used the rotation curves above to estimate the local escape velocity. We follow a straightforward approach in order to build the escape velocity maps. We assume the simplest approximation of the galactic potential, a spherically symmetric one.  From this potential, the circular velocity as a function of galactocentric distance is described by two terms defined by the turn-over radius ($R_\mathrm{turn}$). We derive the escape velocity at a given radius $r$ from the optical nucleus using the following equation:
\begin{equation}
V_\mathrm{esc}^2 (r) = 
\begin{cases}
V_\mathrm{esc,in}^2 (r) + V_\mathrm{esc,out}^2 , & \text{if $r<R_\mathrm{turn}$}\\
V_\mathrm{esc,out}^2 , & \text{if $r>R_\mathrm{turn}$}\\
\end{cases}
\label{eq:v_esc}
\end{equation}

where 

\begin{equation*}
V_\mathrm{esc,in}^2(r) = (V_\mathrm{max}/R_\mathrm{turn})^2 (R_\mathrm{turn}^2 - r^2)
\end{equation*}
and 
\begin{equation*}
V_\mathrm{esc,out}^2 = 2\,V_\mathrm{max}^2\,\ln(R_\mathrm{vir}/R_\mathrm{turn}) + 2\,V_\mathrm{max}^2,
\end{equation*}
where the virial radius ($R_\mathrm{vir}$) is obtained from its relation to the halo mass. The halo mass is in turn obtained by the matching function between halo and stellar mass \citep{Behroozi_2010}.  In the right panel of Fig.~\ref{fig:Example_kin} we plot an example of the profile of $V_\mathrm{esc}$ for a particular galaxy. As expected, the escape velocity decreases with the de-projected distance. In Appendix~\ref{app:EscapeVel}, we explore a more realistic model assuming the baryonic mass is distributed in a disk-like profile and add a spherical NFW dark matter halo. We show that the differences in the escape velocity between this two-component model and the simplified one presented in this section is of the order of $\sim$5\% for a subset of the galaxies included in our sample.   

In Fig.~\ref{fig:Vesc_Sigma} we compare $V_\mathrm{esc}$ against $\Sigma_{*}$,. We find a broad distribution of  $V_\mathrm{esc}$ for a given stellar density. However, the density of the distribution suggests a linear relation (in logarithmic scales) between the escape velocity and the stellar mass density in which the scaling factor depends on the total stellar mass. This is expected given that these observables are tightly related \citep[e.g.,][]{Lelli_2016}. As we note in Sec.~\ref{sec:HaVel}, to fully account for the galaxy potential from a kinematic point of view, in particular for low-mass galaxies, it is necessary to consider the velocity dispersion distribution. For simplicity, in this study we only consider the contribution of the circular velocity. In a future work, we will consider the possible relation of the velocity dispersion in the chemical enrichment at local scales.

\section{Results}
\label{sec:Results}

\subsection{Gas Fraction Metallicity ($\mu$-Z) relation}
\label{sec:OH_fgas}
%
\begin{figure*}
\includegraphics[width=\linewidth]{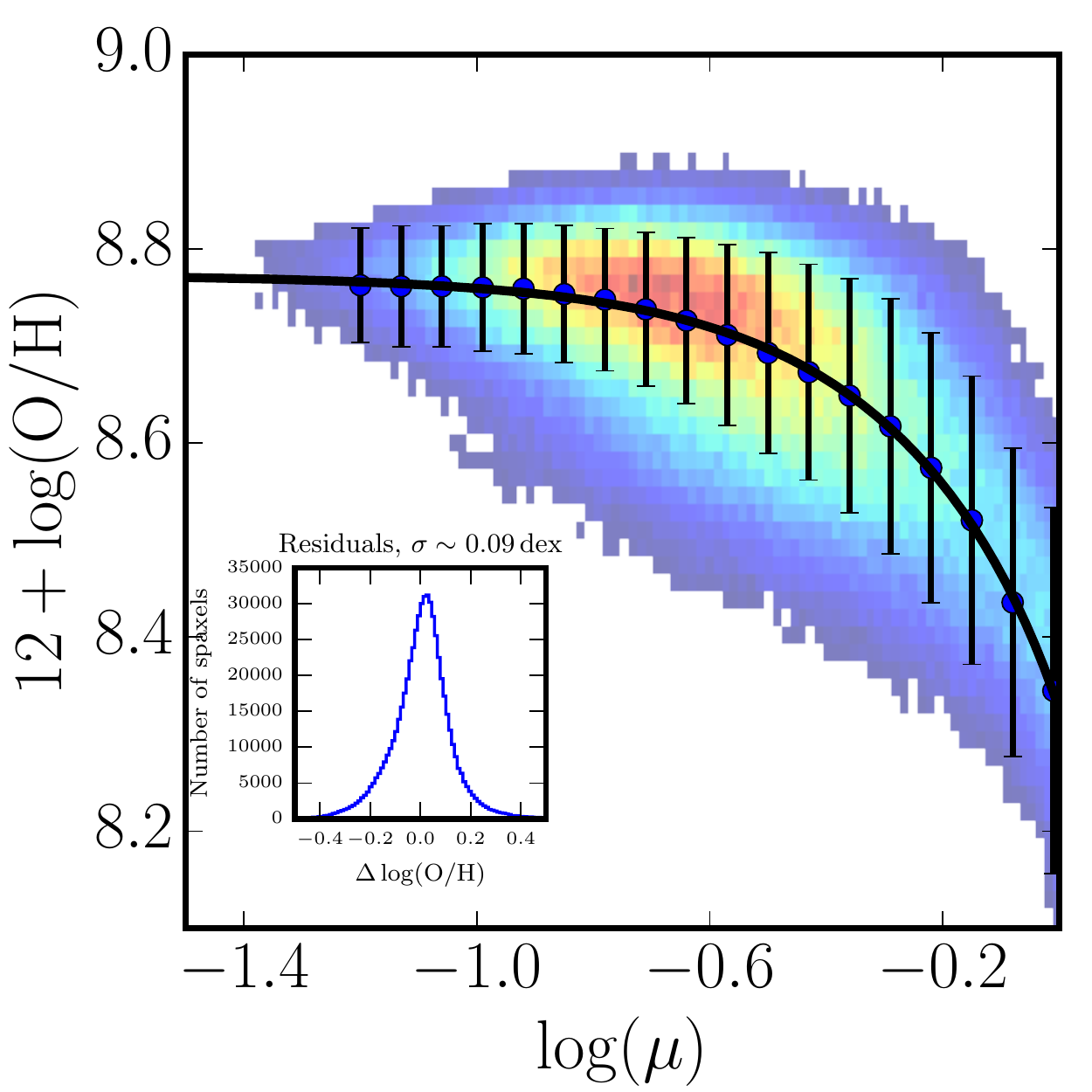}
\caption{Distribution of the metallicity versus the gas fraction ($\mu$) for our sample of galaxies with star forming spaxels ($\sim 9.2 \times 10^5$ spaxels). The colors represent the density of spaxels at different points of the distribution. Blue circles with error bars represent the medians and standard deviations for different bins of the gas fraction, respectively. The solid curve shows the best fitting for the median metallicities using a fourth-order polynomial function. The standard deviation of the residuals from this fitting is of the same order as the residuals of the $\Sigma_*$Z relation ($\sim$ 0.09 dex, see inset).} 
\label{fig:muZ}
\end{figure*}
 
In Fig.\ref{fig:muZ} we plot the distribution of the oxygen abundance as a function of the gas fraction for the star forming spaxels in our sample of 1024 galaxies. We find a very tight relation between these two local observables; as $\mu$ decreases the metal content increases, reaching a constant abundance ($12+\log(\mathrm{O/H})) \sim$ 8.8) for low $\mu$ ($\mu \leqslant$0.1). We measure the median metallicity at different gas fraction bins in the range of $-1.3 < \log(\mu) < -0.1$ within bins of $\log(\mu) \sim$ 0.1. The standard deviation of the metallicity in these bins ranges between 0.03 and 0.1 dex for low and large gas fractions, respectively.  In order to estimate the residuals of this relation, we fit to the median values (blue circles with error bars in Fig.\ref{fig:muZ}) a fourth-order polynomial function.

The median residual of this relation is close to zero ($\sim$ -0.01 dex, see distribution on the inset in Fig.~\ref{fig:muZ}). The standard deviation of these residuals from this best-fitted curve confirms that this is a tight relation ($\sigma \sim$ 0.09 dex). This deviation is comparable to the one observed in the residuals of the $\Sigma_*$-Z relation for the MaNGA galaxies ($\sigma \sim$ 0.06 dex, \cite{BB_2016}). Fig.~\ref{fig:muZ} also shows that for a given intermediate $\mu$,  the metallicity seems to be skewed towards low values. In Sec.~\ref{sec:OH_Vesc} we explore how the residuals of this distribution correlate with the escape velocity. We find that regions with low metallicities tend to have small escape velocities.

As we mentioned above, previous studies aimed at understanding the observed metallicity gradients in small samples of star-forming galaxies \citep{Carton_2015,Ho_2015} found similar trends to those presented in our much larger sample. Despite the tightness of the local $\mu$-Z relation for our large sample of galaxies, we must emphasize that we do not have direct estimations of the molecular or the atomic gas densities. Nevertheless, we consider that this relation is robust regardless of the proxy used to gauge the gas density. In particular, \cite{Carton_2015} used a different method to estimate gas fractions, but found similar results to us. They also reported their derived local $\mu$-Z relation is similar for the H$_I$-rich galaxies and for their control sample, suggesting that the physical processes occurring in both samples are similar. The statistical properties of our sample ensure that we cover a wide range of global parameters (e.g., stellar mass and global SFRs). This, along with the fact that the local $\mu$-Z relation reported here is almost as tight as the $\Sigma_*$-Z relation, indicates that $\mu$ is a key parameter in understanding the metal content of the ionized gas (as noted by \cite{2016A&A...595A..48B}). 

\subsection{Escape Velocity and Metallicity ($V_\mathrm{esc}$-Z relation)}
\label{sec:OH_Vesc}
%
\begin{figure}
\includegraphics[width=\linewidth]{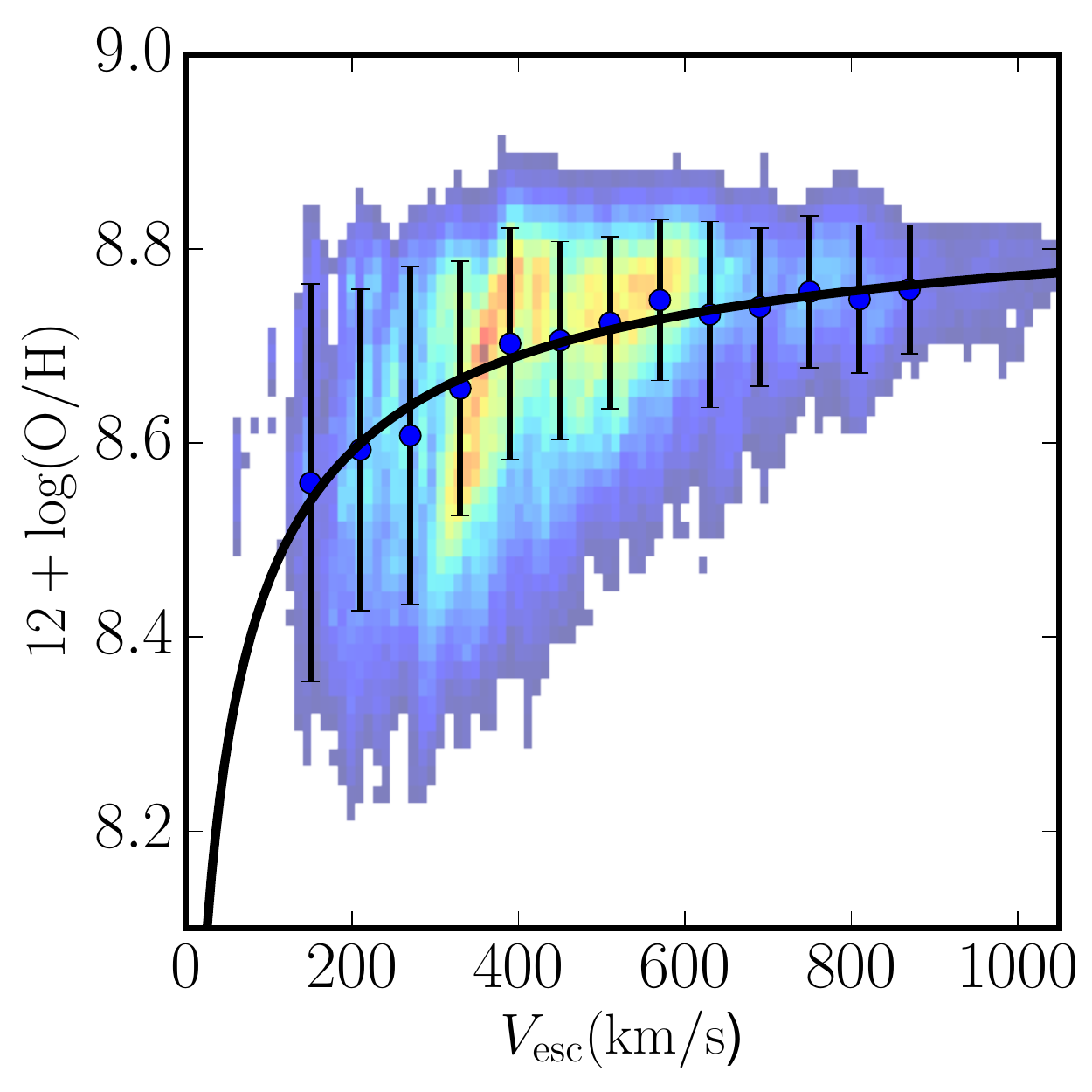}
\caption{Distribution of the local metallicity with respect to the local escape velocity ($V_\mathrm{esc}$). As in Fig.~\ref{fig:muZ}, the colors represent the density of star forming spaxels in the plot with red indicating denser regions. The blue circles with error bars, represent the median metallicities and standard deviations at different bins in $V_\mathrm{esc}$ of 50 km s$^{-1}$ width, respectively. The solid curve shows the best fitting of the median metallicities to the curve describe by Eq.~\ref{eq:fit_VescZ}. The standard deviation of the residuals is slightly larger than the one obtained for the $\mu$Z relation ($\sigma \sim$ 0.11 dex).} 
\label{fig:VescZ}
\end{figure}

The escape velocity parametrizes the amount of material that can be ejected from the galaxy by momentum- or energy-driven outflows. Therefore, it could play a key role in determining the local metallicity. 

In Fig.\ref{fig:VescZ} we plot the distribution of the metallicity as function of the local escape velocity for all the star-forming spaxels in our sample of galaxies. We measure the median metallicity at different escape velocity bins of 60 km/s width, within a range of $ 180\,\mathrm{km\,s^{-1}} < V_\mathrm{esc} < 900\,\mathrm{km\,s^{-1}}$. We find that metallicity increases as $V_\mathrm{esc}$ increases, reaching a constant value of $\sim$ 8.8 dex for  $V_\mathrm{esc} \gtrsim$ 400 km/s. However, the standard deviation for these bins are relatively larger than those obtained for the $\mu$-Z relation ranging from 0.13 dex to 0.04 dex for small and large escape velocities, respectively. We fit to the median metallicities a curve of the form 
\begin{equation}
12 + \log(O/H) = \frac{\alpha} {1+(\beta/V_\mathrm{esc})^\gamma}
\label{eq:fit_VescZ}
\end{equation}
The best fitting values for our data are $\alpha$ = 8.85 $\pm$ 0.3 dex, $\beta$ =  165 $\pm$ 13 km s$^{-1}$, and $\gamma$  = 0.83 $\pm$ 0.3 (see the black solid curve in Fig.~\ref{fig:VescZ}). The median values of the residuals are close to zero ($\sim$ -0.01 dex). The standard deviation of the residuals from this best-fit curve are slightly larger than the ones derived from the $\mu$Z relation ($\sigma \sim$ 0.11 dex). The distribution of these two parameters is not symmetrical with respect to the best-fit line. In particular, at low values of  $V_\mathrm{esc}$, the metallicity has a tail skewed towards lower values. We find that these low-metallicity spaxels have large gas fractions. 

Bearing in mind these results, in the next section we will investigate the impact of the escape velocity on the $\mu$-Z relation (and vice versa).  

\subsection{Joint Dependences on $V_\mathrm{esc}$ and $\mu$}
\label{sec:OH_Vesc}
%
\begin{figure}
\includegraphics[width=\linewidth]{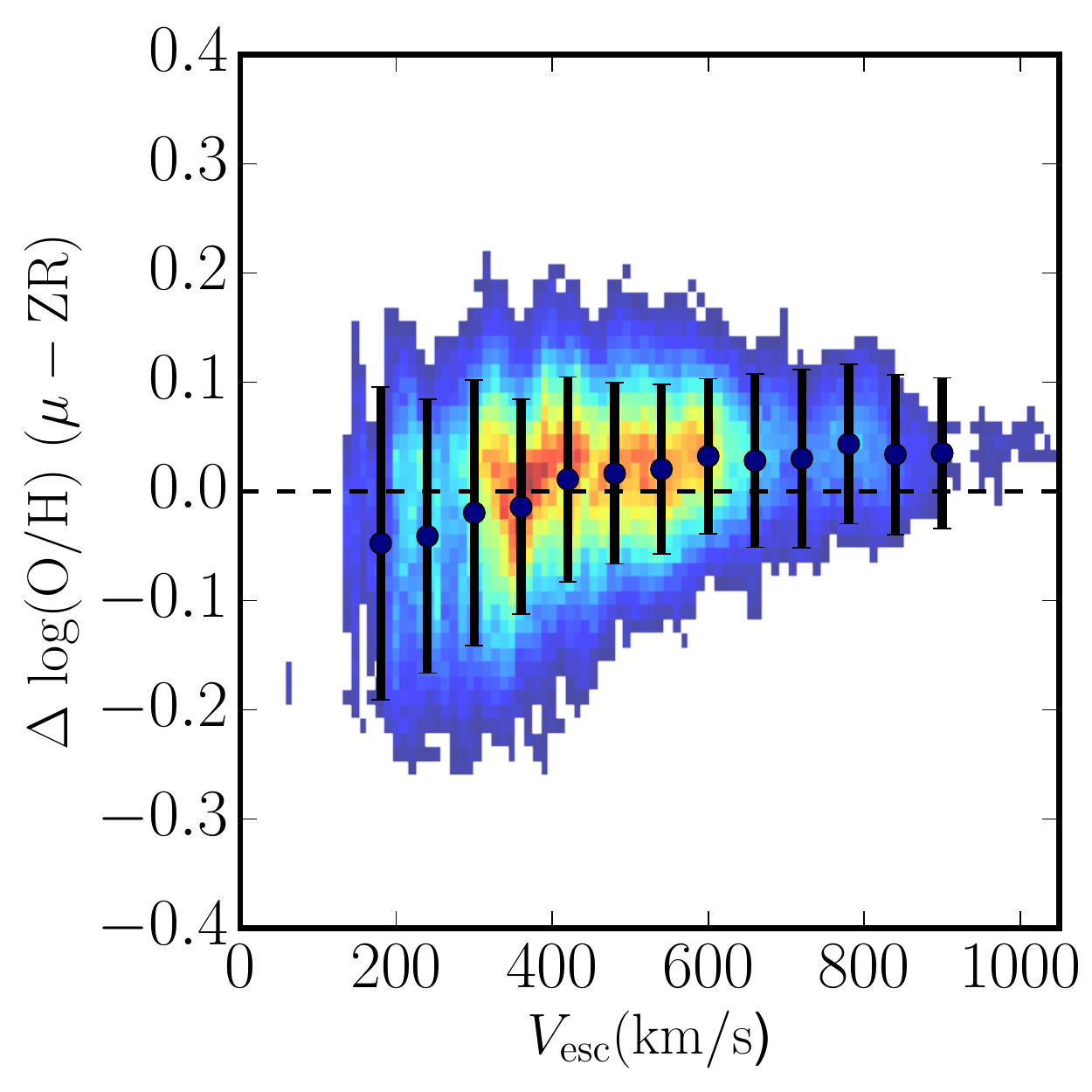}
\caption{Distribution of the residuals from the $\mu$-Z relation versus the escape velocity $V_\mathrm{esc}$. As in the previous figures the color represents the density of star forming spaxels in the distribution, with red indicating higher density. The circles and error bars represent the median and standard deviation of the residuals in different bins of $V_\mathrm{esc}$ of 50 km/s width each, respectively. The dashed lines represent the best-fit curve in the $\mu$-Z relation (i.e., zero scatter).} 
\label{fig:dmuZ_Vesc}
\end{figure}
%
\begin{figure}
\includegraphics[width=\linewidth]{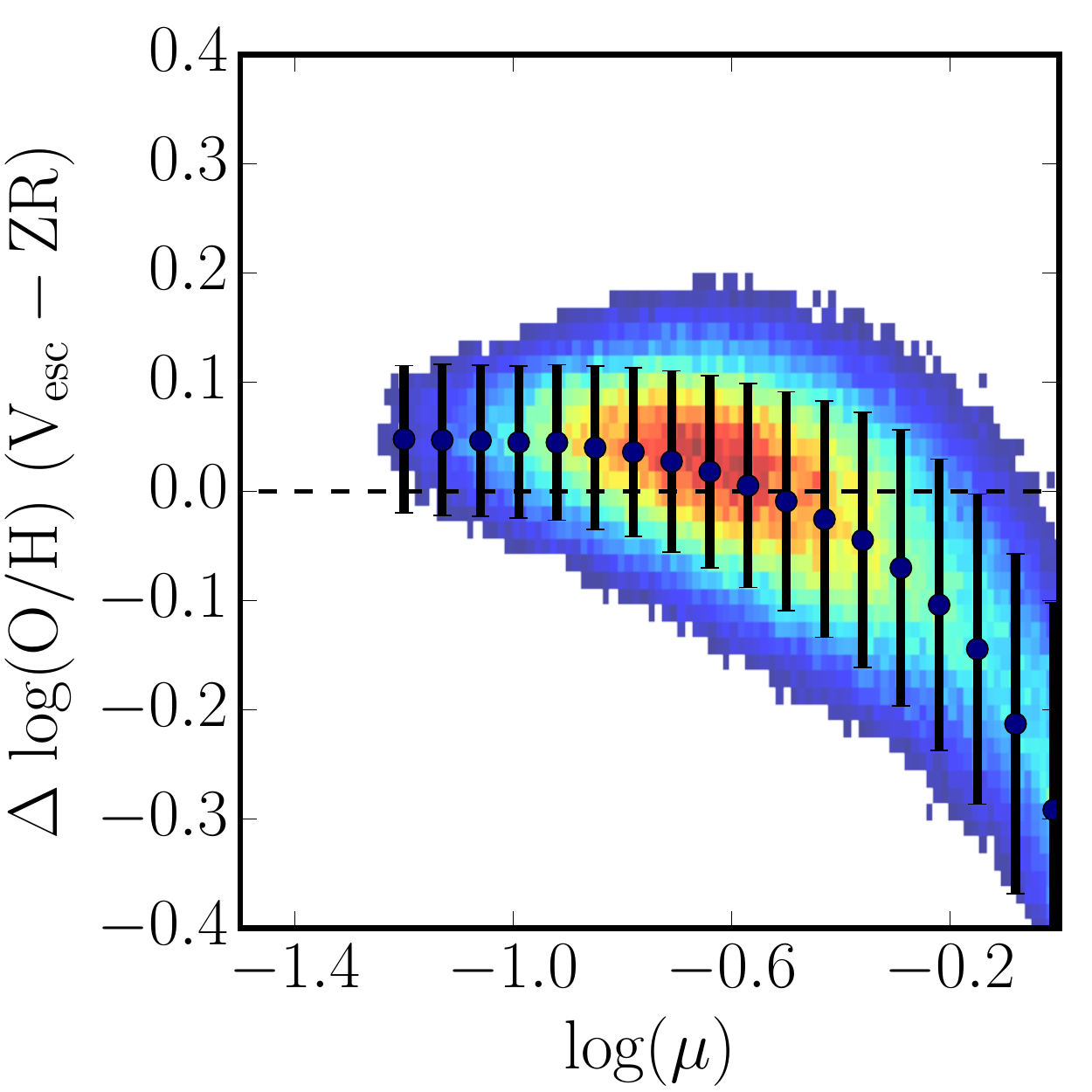}
\caption{Distribution of the residuals from the  $V_\mathrm{esc}$-Z relation versus the gas fraction. As in previous figures the color represents the density of star forming spaxels in the distribution, with red indicating higher density. The circles and error bars represent the median and standard deviation of the residuals in different bins of $\log(\mu)$ of 0.1 dex width, respectively. The dashed lines represent the best fitted curve in the $\mu$-Z relation (i.e., zero scatter).} 
\label{fig:dVescZ_mu}
\end{figure}

So far we have investigated how the local metallicity relates to the gas fraction and  $V_\mathrm{esc}$ separately. In this section we would like to explore how $\mu$ and $V_\mathrm{esc}$ can jointly regulate the local metal content in our sample of galaxies. To explore this joint dependence, we first study the residuals of the $\mu$-Z relation presented in Sec~\ref{sec:OH_fgas} as function of $V_\mathrm{esc}$. We then study how the residuals of the relation of the metallicity with $V_\mathrm{esc}$ relates to the gas fraction.

In Fig.~\ref{fig:dmuZ_Vesc} we plot the distribution of the residuals from the $\mu$Z relation as function of $V_\mathrm{esc}$. Broadly speaking, there is a mild trend for residuals to increase with $V_\mathrm{esc}$. Using the median metallicities at different $V_\mathrm{esc}$ bins, we find that negative residuals (i.e., overestimation of the best fitting curve, by $\sim$ 0.04 dex) are found in spaxels with $V_\mathrm{esc} \lesssim$ 400 km/s. For larger $V_\mathrm{esc}$ the residuals seem to have a constant positive value (i.e., underestimation of the best fitting curve) of $\Delta \log(\mathrm{O/H}) \sim$ 0.03 dex. We also note that the scatter of these residuals tend to increase as $V_\mathrm{esc}$ decreases. In Sec.~\ref{sec:Models} we study the impact of parameterizing the expelled gas as a function of $V_\mathrm{esc}$ in models of chemical evolution.

In Fig.~\ref{fig:dVescZ_mu} we study the residuals of the $V_\mathrm{esc}$-Z relation against the gas fraction. We find a clear trend of negative residuals increasing at larger gas fractions, with an amplitude of about 0.3 dex.  This implies that those spaxels where the fitted $V_\mathrm{esc}$-Z curve overestimate the observed metallicity tend to be those with larger gas fractions. 

In summary, we find that the primary factor in determining the local metallicity is the local gas fraction, with the local escape velocity playing a smaller role.  Bearing these results in mind, in the next section we will compare the observed $\mu$-Z relation with several simple analytic models of galactic chemical evolution. We will show that in order to explain the observed local metallicities, it is required to take into account the fraction of metals removed from a local region, as parametrized by $V_\mathrm{esc}$. 

\section{Analitical Modelling of the local Metallicity}
\label{sec:Models}

In this section we will analyze two recent models of chemical evolution in order to gain insight into the most plausible physical scenario for the evolution of metals at local scales in star forming galaxies. More specifically, we will compare our MaNGA observed $\mu$-Z relation with the predictions from these analytic models. First, we briefly explain the gas-regulator model proposed by \cite{Lilly_2013} and implemented at local scales by \cite{Carton_2015} (Sec.~\ref{sec:GasReg}). We will then also summarize the local leaky-box model proposed by \cite{Zhu_2017} (Sec.~\ref{sec:LeakyBox}). For both models we parametrize outflows as a function of the local escape velocity (Sec.~\ref{sec:outflow_par}). The results of this analysis are presented in Sec. ~\ref{sec:res}.

%
\subsection{The gas-regulator model}
\label{sec:GasReg}
%

The main idea in the model of chemical evolution presented by \cite{Lilly_2013}, relies on the postulate that the global star formation rate (SFR) is regulated by the total mass of gas (also known as the gas reservoir) in the galaxy. In turn, any other process that affects the fraction of gas or metals in the ISM (i.e., gas transformed into long lived stars, metal-rich material returned instantaneously from the stellar component, and gas expelled from the system) can be represented as function of the SFR.

More recently, \cite{Carton_2015} adapted this model to explain the metallicity gradients in a sample of star-forming galaxies. For this local adaptation of the global model, they apply the model in radial bins (annuli). Then, for each of these bins they assume an equilibrium between inflows, outflows and star formation. They also assume that radial transfer of gas/stars/metals between the bins could be ignored. A detailed description of the spatially resolved gas-regulator model is given in \cite{Carton_2015}, whereas a full description of the global model is found in \cite{Lilly_2013}. Here we briefly outline the main features of the gas regulator model presented in \cite{Carton_2015}. We use the similar assumptions as in \cite{Carton_2015} implement for the radial bins in their sample of galaxies. 


The change in time of the gas mass in each of the spaxels is then given by

\begin{equation}
\dot{m}_\mathrm{gas} = \dot{m}_\mathrm{in} - \dot{m}_\mathrm{out} - \dot{m}_\ast + \dot{m}_\mathrm{return}.
\label{eq:reg_gas_basic}
\end{equation}
where, $\dot{m}_\mathrm{in}$ represents the inflow of metal-poor gas into the spaxel and $\dot{m}_\mathrm{out}$ is the rate at which enriched material leaves the spaxel. \cite{Carton_2015} assumed that this was due to outflows driven by from massive stars and core-collapse supernovae. They therefore considered this to be proportional to the SFR by $\dot{m}_\mathrm{out}$ = $\lambda$ SFR, where $\lambda$ is the mass loading factor (see below Sec.~\ref{sec:outflow_par}). 

As we mentioned above, this model is based on the fact that the SFR is proportional to, regulated by the gas mass. Thus, SFR = $\epsilon$ $m_\mathrm{gas}$, where $\epsilon$ is the star formation efficiency. Finally, $\dot{m}_\mathrm{return}$, represents the rate at which metal enriched gas returns to the ISM from short-lived high mass stars. The model assumes that a fraction $R$ of the mass converted into stars returns instantaneously to the gas reservoir. Thus, following \cite{Lilly_2013}, and  \cite{Carton_2015} we assume that this fraction is constant ($R$ = 0.4). In simple terms, Eq.~\ref{eq:reg_gas_basic} is a representation of mass conservation: the fraction of gas that enters the system could end up as long lived-stars, be expelled by the system, or be recycled \citep[see Fig.~2 in ][]{Lilly_2013}. With these assumptions the change of the gas in the reservoir is given by
\begin{equation}
\dot{m}_\mathrm{gas} = \dot{m}_\mathrm{in} - (1 - R + \lambda) \textrm{SFR} .
\label{eq:reg_gas}
\end{equation}

By defining the gas-to-stellar mass ratio as $r_\mathrm{gas} = m_\mathrm{gas} / m_\ast$, \cite{Lilly_2013} show that the above equation can be expressed as 
\begin{equation}
\dot{m}_\mathrm{in} = \left( (1 - R) (1 + r_\mathrm{gas}) + \lambda + \epsilon^{-1} \frac{d\ln(r_\mathrm{gas})}{dt}\right) \cdot \textrm{SFR},
\label{eq:reg_gas_rearranged}
\end{equation}
This equation explicitly relates the inflow rate to the star formation rate. We assume this relation is valid for our sample of star forming spaxels. The gas-to-stellar mass ratio can be easily transformed to the gas fraction by
\begin{equation}
r_\mathrm{gas} = \frac{\mu}{1-\mu}
\label{eq:r2mu}
\end{equation}
Similar to Eq.~\ref{eq:reg_gas_basic}, the change of metals in the reservoir is related to the inflow of metal-poor gas from the halo, the recycled metal-rich material produced by short-lived stars, the material transformed into long-lived stars and the enriched material ejected out of the reservoir via outflows. Along with these assumptions and Eq.~\ref{eq:reg_gas_rearranged}, \cite{Lilly_2013} find that this gas-regulator system will reach equilibrium on time-scales shorter than the depletion time-scale (i.e., $\leqslant$ 1/$\epsilon$). They show that in equilibrium the metallicity of the gas-regulated system can be described as
\begin{equation}
Z_\mathrm{eq} = Z_{0} + \frac{y}{1 + r_\mathrm{gas} + (1-R)^{-1} \left(\lambda + \epsilon^{-1} \frac{d\ln(r_\mathrm{gas})}{dt}\right)}.
\label{eq:reg_equib_metal}
\end{equation}
where $Z_{0}$ and $y$ are the infalling Oxygen mass fraction and yield (i.e., the Oxygen mass returned per unit mass in long-lived stars), respectively. This equation explicitly correlates the metallicity with the gas fraction $r_\mathrm{gas}$.
The fitting of this model to SDSS data by \cite{Lilly_2013} results in  $\epsilon^{-1} \frac{d\ln(r_\mathrm{gas})}{dt}$ $\sim$ -0.25. We set this factor to that constant value.
\subsection{The Leaky-Box Model}
\label{sec:LeakyBox}
%

Recently, \cite{Zhu_2017} examined the local $\Sigma_\ast$-Z relation observed in MaNGA galaxies \citep{BB_2016}. They assume an evolutionary model in which disk galaxies grow inside out, with metal-poor gas accreting from the halo to the outskirts of the galaxy collapsing and eventually triggering localized star formation. They assume that there are no radial flows in the disk plane. Hence, for a given region within a disk galaxy they define a total density $\Sigma_{\rm tot}(t)$ that does not vary with time and is set only by the accreted initial gas surface density at time $t_0$ $\Sigma_0$ =  $\Sigma_{\rm gas}(t_0)$. The evolution of the localized components is then given by 
\begin{eqnarray}
\Sigma_{\rm tot}(t) & = & \Sigma_*(t) + \Sigma_{\rm gas}(t) + \Sigma_{\rm out}(t) \\
& = & \Sigma_0\,\mathrm{,}  \nonumber 
\label{eq:sigma0}
\end{eqnarray}
where $\Sigma_{\rm out} (t)$ represents the density of the expelled gas that does not return to the galaxy. Similar to \cite{Lilly_2013}, the outflow rate is related to the SFR surface density by the mass loading factor. Under these assumptions the  evolution of the metallicity is 
\begin{equation}
Z = Z_0+\frac{y}{1+\lambda} \ln \left( \frac{\Sigma_0}{\Sigma_{\rm gas}} \right)
\label{eq:metalgas}
\end{equation}

where $y$ is the effective yield and $\Sigma_0 = \Sigma_{\rm gas} + (1+\lambda) \Sigma_{*}$. Rewriting this in terms of the total gas fraction,  the observed metallicity can be described as:
\begin{equation}
Z = Z_0+\frac{\ln(10)y}{1+\lambda} \cdot \log\left( 1 + \frac{(1+\lambda)(1-\mu)}{\mu} \right) \,\mathrm{.}
\label{eq:metalgas}
\end{equation}
As in the case of the gas regulator model, now we have an explicit relation between the expected metallicity and the gas fraction $\mu$. We note that \cite{Zhu_2017} assumed a constant value for $\lambda$. We will relax this assumption below and introduce a dependence on $V_\mathrm{esc}$.

\subsection{Local parametrization of the outflows}
\label{sec:outflow_par}

As we mentioned above, both models described in this study correlate the gas/metal loss from outflows with the SFR via the mass loading factor ($\lambda$). Observations of starburst galaxies reveal that galactic-scale outflows have properties that indeed relate to star-formation rate and star-formation rate per unit area \citep[]{Heckman_2000, Rupke_2002, Martin_2005, Grimes_2009, Hill_2014, Heckman_2015, Heckman_2016}. The measured outflow velocities ($v_w$) are of the order of hundreds of km s$^{-1}$. Simple analytic models assume that these outflows are driven either by momentum or energy. Assuming that outflows are driving by a combination of hot wind fluid driven\citep[thermalized ejecta of massive stars and supernovae][]{Chevalier_1985} and radiation pressure \citep{Murray_2005}. \cite{Heckman_2015} show that for a typical starburst population the total momentum flux is given by $\dot{p}_* = 4.8\times10^{33}\mathrm{\,\,SFR}$ dynes. For a momentum-driven outflow  ($\dot{p}_* = \dot{M}_* v_{out}$), the mass loading factor is then given by
\begin{equation}
\lambda_\mathrm{momentum} = \frac{670\,\,\mathrm{km\,\,s^{-1}}}{v_\mathrm{w}}.
\end{equation}

On the other hand if the outflow is driven by kinetic energy from supernovae and stellar winds \citep{Chevalier_1985,Dekel_1986,Silk_1998,Murray_2005}, the mass loading factor is described by

\begin{equation}
\lambda_\mathrm{energy} = \frac{2 \epsilon_\mathrm{eff} \eta_\mathrm{SN} \mathrm{E_{SN}}}{ v_w^2 } = \left(\frac{1000\,\,\mathrm{km\,\,s^{-1}}}{v_\mathrm{w}} \right)^2
\end{equation}
where $\mathrm{E_{SN}} = 10^{51} \mathrm{erg}$ is the typical energy produced by a SN,  
 $\eta_\mathrm{SN} = 1 \times 10^{-2}$, is the number of SNe per solar mass, and $\epsilon_\mathrm{eff}= 1$ is the efficiency in which supernova transfer kinetic energy to the ISM. In Sec.~\ref{sec:dis_models} we compare the above normalizations of the mass loading factor with the results from fitting the models to the local relation. To fit the models presented in previous sections with the data, we use a local version of the parametrization of the mass loading factor for global parameters proposed by \cite{Peeples_2011}
\begin{equation}
\lambda(r) = \left(\frac{V_0\,\,[\mathrm{km\,\,s^{-1}}]}{V_\mathrm{esc}(r)} \right)^\alpha + \lambda_0
\label{eq:lam_par}
\end{equation}
in the next section we will study the impact in the fitting using different selections of the values that parametrize the mass loading factor. 
\subsection{Fitting the models to the data}
\label{sec:res}
\begin{figure*}
\includegraphics[width=\linewidth]{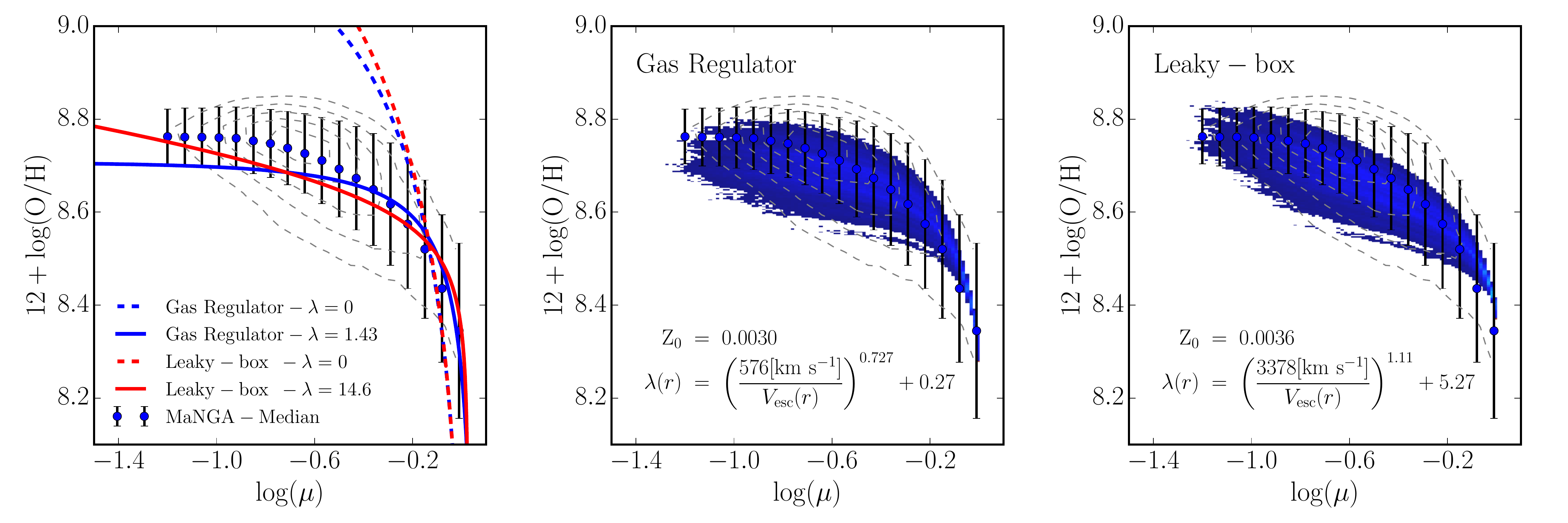}
\caption{Comparison between the predicted metallicity from models of local chemical evolution and the observed MaNGA $\mu$-Z relation. In all panels the blue dots with error bars are the same as in Fig.~\ref{fig:muZ}. The outermost and innermost gray dashed contours represent the  90\% and 10\% of the distribution presented in Fig.~\ref{fig:muZ}, respectively. Left panel: The solid and dashed blue lines represent the metallicity expected by a gas regulator model with $\lambda(r) = 0$ and $\lambda(r) = 1.4$ (see details in Sec.~\ref{sec:GasReg}), whereas the solid and dashed red lines show the metallicity expected from a leaky-box model with $\lambda(r) = 0$ and $\lambda(r) = 14.6$ (see details in Sec.~\ref{sec:LeakyBox}). For both models a windless scenario results in similar $Z_0 \sim 6\times10^{-4}$, whereas a those in which a constant $\lambda$ was fitted results in $Z_0 \sim 3\times10^{-3}$ (roughly half solar Oxygen abundance). The middle and right panels show the best fits of the models (gas regulator and leaky box, respectively) when we parametrize the mass loading factor by the local escape velocity (see Eq.~\ref{eq:lam_par}). For both models we fix the oxygen yield to $y = 0.014$.} 
\label{fig:Model_fgas}
\end{figure*}

In both models described in Secs.\ref{sec:GasReg} and \ref{sec:LeakyBox} there are three unconstrained parameters, the oxygen yield $y$ (mass of newly created oxygen returned to the ISM, divided by the mass of long lived stars), the metallicity (oxygen mass fraction) of the inflowing material $Z_0$, and the mass loading factor $\lambda(V_{\rm esc})$. 

As \cite{Peeples_2011} point out, variations in the IMF as well as uncertainties in Type II supernova yields lead to a rather poorly constrained yield ($0.008 \leqslant y \leqslant 0.023$). Nevertheless, these authors justify a mid-range value of  $y = 0.015$ by comparing different nucleosynthetic yields and IMFs (see their Fig. 7). We find similar values by studying the oxygen mass loss in a synthetic galaxy modeled using \texttt{STARBURST99} \citep[y = 0.014, ][]{Leitherer_2014}. A simple empirical way to constrain the Oxygen yield is to use clusters of galaxies, assuming that they are closed boxes \citep{Renzini_2014}. In this case, nearly all the Oxygen produced will either be in stars or the hot intra-cluster medium (ICM).  We assume that Oxygen in the cold ISM is negligible in these early-type gas-poor galaxies. We can then write the following expression for the total Oxygen mass
\begin{equation}
M_O =  <Z_*> M_* +  <Z_{\rm ICM}> M_{\rm ICM}
\end{equation}
where $<Z_*>$ and $<Z_{\rm ICM}>$ are the mean mass fractions of Oxygen in the stars and ICM respectively. Since the stellar mass in the early-type galaxies that dominate clusters is almost entirely made-up of long-lived stars and remnants, we will use the present day M$_*$ to define the yield. That is,
 
\begin{equation}
Y_O = M_O/M_* = <Z_*> + <Z_{\rm ICM}>(M_{\rm ICM}/M_*)
\end{equation}
\cite{Renzini_2014} take a mean stellar metallicity of solar, and estimate that $M_{\rm ICM}/M_*$ = 5.6. The radial metallicity gradients in the ICM in \cite{Mernier_2017} and \cite{Simionescu_2015} imply a mean metallicity of $\alpha$-elements about 0.25 solar.  The implied Oxygen yield is then 0.006 + 0.0015 $\times$ 5.6 = 0.0144. This is gratifyingly close to the estimate from \texttt{STARBURST99}, and we therefore fix the yield in our models to 0.014. With this, we select three different variations of the mass loading factor in our models: a windless scenario ($ \lambda = 0$), a constant positive loading factor ($ \lambda = \mathrm{constant} > 0$) and a variable mass-loading factor that varies with $V_\mathrm{esc}$ as in Eq.\ref{eq:lam_par}. For all fits, we set $Z_0$ as a free parameter. 

In Fig.\ref{fig:Model_fgas}, we present the results of the fitting, allowing different parametrizations as explained above. The left panel shows the results of the fitting for both models assuming a windless scenario ($\lambda = 0$, dashed lines) and fitting the mass loading factor as a constant value (solid lines). For both models in the windless scenario the metallicity of the accreting gas is $Z_0 \sim 6\times10^{-4}$, while in the constant mass loading factor models $Z_0 \sim 3\times10^{-3}$. This latter value is about half the solar oxygen abundance. We will comment on this below. 

From this panel it is evident that a windless scenario provides a very poor fit to the observed $\mu$-Z relation. On the otherhand when a constant mass loading factor is included in both models, the fit to the observed metallicities is significantly improved. However, the value derived for the leaky box model ($\lambda \sim 14 $) is one order of magnitude larger than the one derived from the gas regulator model ($\lambda \sim 1.4$). As we will discuss in Sec.~\ref{sec:Discu}, the large mass-loading factor in the leaky-box model is not compatible with other observed properties of galaxies.

In middle and right panels of Fig.~\ref{fig:Model_fgas} we compare the observed $\mu$-Z relation with the gas regulator  and leaky box best fits from models parameterizing the local mass loading factor as in Eq.~\ref{eq:lam_par}, respectively. For both models the best fit is no longer a single line, but a two dimensional distribution. This is a consequence of assuming the dependence of the mass loading factor on the local  $V_\mathrm{esc}(r)$. For both models the metallicity of the accreting gas is $Z_0 \sim 3\times10^{-3}$). Once again, the fitted values in the parametrization of the mass loading factor using Eq.~\ref{eq:lam_par} are much larger in the leaky-box model in comparison to the gas regulator model.

\section{Discussion}
\label{sec:Discu}
%
%
In Sec.~\ref{sec:Analysis} we presented the relation between the metallicity and the total gas fraction ($\mu$, see Fig.~\ref{fig:muZ}). We showed that the dispersions in the residuals are similar to those observed in the local stellar surface mass density relation ($\Sigma_*$-Z) reported for a similar sample of MaNGA galaxies \citep{BB_2016}. Thanks to the wealth of information in the MaNGA data cubes, we are also able to make spatially-resolved estimates of the escape velocity. We found a relation between metallicity and $V_\mathrm{esc}$ (see Fig.\ref{fig:VescZ}). The residuals of these two relations indicate the interplay of $\mu$ and $V_\mathrm{esc}$ in producing the observed metallicity (see Figs.~\ref{fig:dmuZ_Vesc} and \ref{fig:dVescZ_mu}). 

From these two plots we inferred that the role of the escape velocity in the local chemical enrichment is secondary in comparison to the gas fraction. However, we also showed that including a parametrization of the mass loading factor that depended on the local escape velocity in models of chemical evolution resulted in a better description of the observed $\mu$-Z relation (see Fig.\ref{fig:Model_fgas}).  In this section we discuss the validity of our results. First, we explore the implications of the best-fitted parameters derived from the two chemical models presented in Sec.~\ref{sec:Models}. Then we discuss the role of the SFR surface density, surface gas density and escape velocity in the $\mu$-Z relation. 

\subsection{Chemical evolution models}
\label{sec:dis_models}

In Sec.~\ref{sec:Models} we compared the derived metallicity from different models of chemical evolution to the observed metallicity in order to understand the physical scenario that explains the local $\mu$-Z relation and the role of $V_\mathrm{esc}$. Specifically, we compared local adaptations of the gas-regulator  (see Sec.~\ref{sec:GasReg}) and the leaky-box models (see Sec.~\ref{sec:LeakyBox}). The gas-regulator model \citep{Lilly_2013}, adapted for local scales by \cite{Carton_2015}, assumes that all the processes able to enrich the ISM (i.e, SFR and flows) are directly related to the gas reservoir in each of the regions (or spaxels) included in the host galaxy (see Eq.~\ref{eq:reg_gas_basic}). \cite{Carton_2015} note that the main parameters that affect the determination of the metallicity gradients in their sample were the local gas fraction, and the mass ejected out of the host galaxy(i.e., the mass-loading factor $\lambda$). The other chemical scenario that we use, the leaky-box model, assumes that the total surface mass density in a given region of a galaxy (plus the mass ejected) is set by the accreted initial gas density ($\Sigma_0$). In this model, the metallicity is constrained mainly by $\lambda$ and $\mu$ (see Eq.~\ref{eq:metalgas}). We note that these are simplified models of the chemical evolution of the ISM that ignore other possible effects that can alter the chemical evolution (e.g., radial flows in the disk).

Our results show that the simple prescription provided by a windless scenario ($\lambda = 0$) at local scales provides a poor fit to the observed metallicity (see dotted lines in left panel of Fig.~\ref{fig:Model_fgas}) in both of the models. In other words, the gas fraction alone and the assumption of instantaneous recycling of metals from long-lived stars to the ISM are not sufficient to explain the observed local metallicity.  On the other hand, a positive constant mass loading factor improved the fit to the observed local $\mu$-Z relation in both models. However the best-fitted parameters we obtain for each model are different ($\lambda \sim $1 and 14 for the gas regulator and leaky box models, respectively). 

The large mass-loading factor of the leaky-box model is not tenable for several reasons. First, from molecular and atomic gas studies in nearby galaxies \citep[e.g., ][]{Leroy_2008} the gas depletion time ($\tau_\mathrm{dep}$), which is defined as the ratio  $\Sigma_\mathrm{gas} / \Sigma_\mathrm{SFR}$ is relatively constant for most nearby galaxies, with a value of $\sim 10^9 \mathrm{yr}$. Assuming that there is no replenishment of accreted gas, for the leaky box model all the observed gas will be consumed and/or expel in $\sim \tau_\mathrm{dep}/15$, or less than 100 Myr. This is implausibly short. Second, the mass-loading factors would require that about ten times more metals are outside of galaxies than are inside. This is inconsistent with observations, as well as the metal mass budget that could have been created in stars  \citep{Peeples_2014}. Finally, the amount of momentum needed to drive the very high mass outflow rates in the leaky-box model exceeds the total available amount by roughly an order-of-magnitude, as we now show.

\begin{figure}
\includegraphics[width=\linewidth]{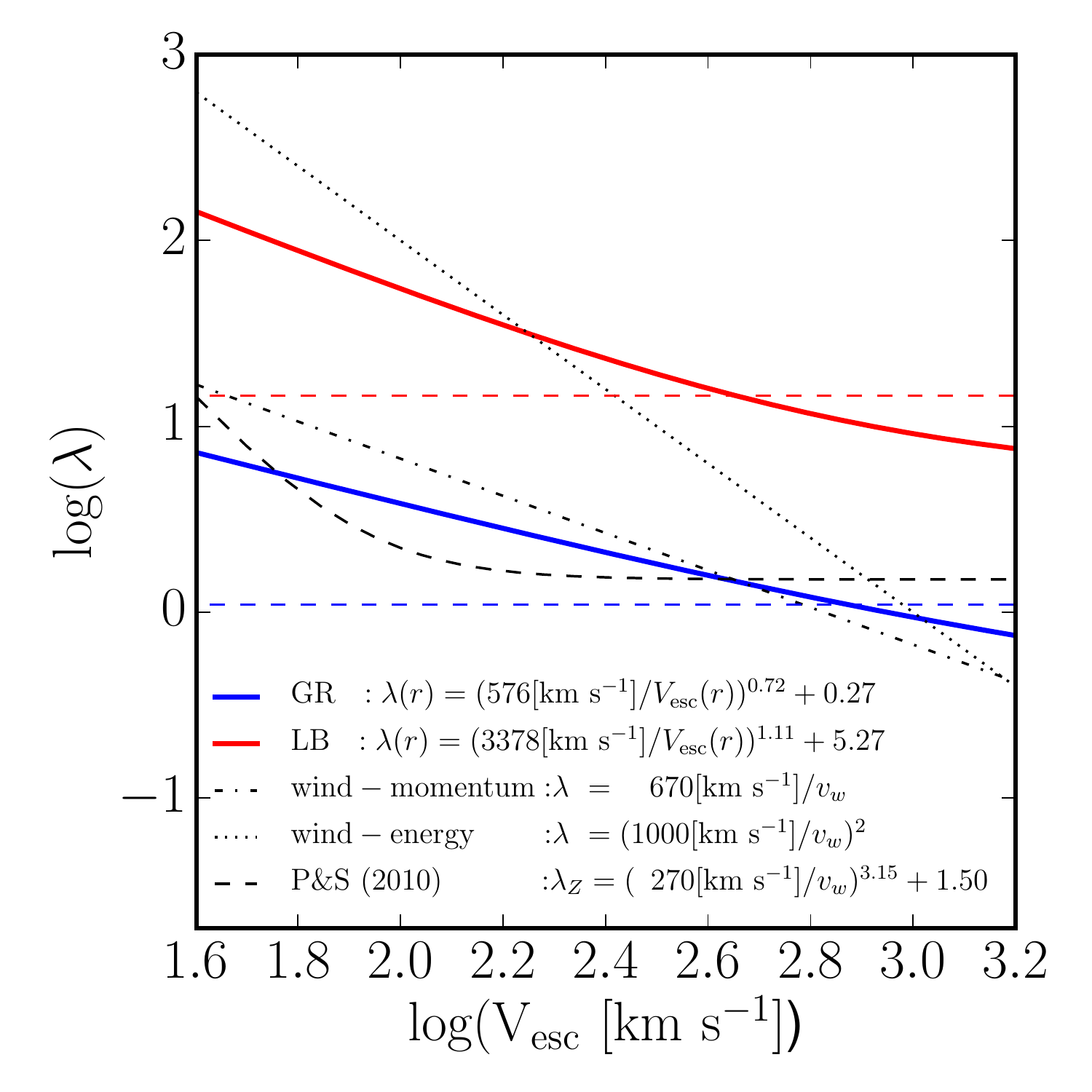}
\caption{Mass loading factor as function of $V_\mathrm{esc}$ using the best-fitting parameters from the modeled local $\mu$-Z relation. Blue and red solid curves represent the gas regulator and leaky-box model, respectively. Blue and red dashed lines represent the best-fitting parameters when assuming a constant loading factor for the gas regulator and leaky-box model, respectively. For comparison purposes, we over-plot the mass loading factor derived for the same metallicity calibrator using integrated (global) properties derived in \cite{Peeples_2011}. } 
\label{fig:lambda_Vesc}
\end{figure}

In Sec.~\ref{sec:outflow_par} we parametrized the loss of enriched gas in terms of the local escape velocity (Eq.~\ref{eq:lam_par}), and fit our model to the data using this parametrization (see middle and right panels of Fig.~\ref{fig:Model_fgas}). We found different sets of best-fitted parameters for the two models of chemical evolution. In Fig.~\ref{fig:lambda_Vesc} we compare the mass loading factor as a function of $V_\mathrm{esc}$ using the best-fitted parameters derived in Sec.~\ref{sec:res} for both the gas regulator and leaky box models (blue and red lines, respectively). As noted above, $\lambda(V_\mathrm{esc})$ derived from the best-fitting of the leaky box model is one order of magnitude larger than the one derived from the gas regulator model. We also compare our best-fitted $\lambda$ for local $V_\mathrm{esc}(r)$ with theoretical models of supernova winds driven by momentum and energy described in Sec.~\ref{sec:outflow_par} (dashed and dotted-dashed lines in Fig.~\ref{fig:lambda_Vesc}). While the leaky-box model yields a best-fit slope consistent with a momentum-driven outflow, the normalization is about an order-of-magnitude too high. On the other hand, the gas-regulator model is in agreement with a momentum-driven outflow in terms of both the normalization and slope. Energy-driven winds predict a large overestimation of $\lambda$ and a steeper slope compared to the best-fits to either the gas-regulator or leaky-box models. Finally, the dashed curve in Fig.~\ref{fig:lambda_Vesc} shows $\lambda$ for the chemical evolution model presented by \cite{Peeples_2011}. We plot the best-fitting parameters derived for the same metallicity calibrator used in this study \citep[i.e., PP04O3N2,][]{2004MNRAS.348L..59P}. For comparison, we assume that $v_w = 3 v_\mathrm{vir}$. Despite the difference between the methods used to derive the escape velocity, scales (local vs global), and parametrization, the local $\lambda (r)$ from the gas regulator model is consistent to the one derived by \cite{Peeples_2011}. 

In the chemical models used in this study, we implicitly assumed that the metallicity of the expelled gas due to outflows is similar to ISM metallicity ($Z_\mathrm{ISM}$). \cite{Peeples_2011} fit a metallicity-weighted mass loading factor ($\lambda _Z = [Z_w /Z_\mathrm{ISM}]\,\lambda$) to the global parameters. In this model the metallicity of the outflow could be larger than the surrounding ISM. Their best fitting parameters yield in general a steeper $\lambda_Z$ in comparison to simple analytical models we have considered. 

However, the results presented here from spatially-resolved data, and assuming a local version of the gas regulator model, suggest that outflows driven by feedback from massive stars are likely to be momentum driven across the disk of the galaxy. 

For simplicity most models of chemical evolution assume the metallicity of the accreted gas to be zero. However, the metallicity of the intergalactic medium $Z_\mathrm{IGM}$ is not zero at the current epoch. There are indications that IGM has been enriched since early epochs \citep[$z > 3$, e.g.,][]{Songaila_1996,Ellison_2000,Schaye_2003}. In the chemical models used in this study we set as a free parameter the metallicity of the accreted material, $Z_0$ (see Sec.~\ref{sec:Models}). Regardless of the parametrization of mass loading factor for both models, we find $Z_0 \sim 3\times10^{-3}$ corresponding to $Z_0 \sim \times10^{-0.30} Z_{\odot}$. Although there are major uncertainties in the amount of metals expelled at high redshift, and in recycling time scales to deliver these back to galaxies, this value is similar to recent estimates of the metallicity of the Circum-Galatic Medium at low redshift: $Z_\mathrm{IGM} \sim 10^{-0.51} Z_{\odot}$ \citep{Prochaska_2017}.   

In conclusion, we used the local adaptation of two chemical models (gas regulator and leaky-box) in order to understand the observed local $\mu$-Z relation derived from the MaNGA survey. From the fitting of these models we find that the gas regulator model provides a better description of the observed parameters. In particular, we find that in this model the best parametrization of the mass loading factor as a function of local escape velocity is in good agreement with models of momentum-driven outflows powered by feedback from massive stars and supernovae.

\subsection{The Role of Star Formation}
\begin{figure*}
\minipage{0.495\textwidth}
\includegraphics[width=\columnwidth]{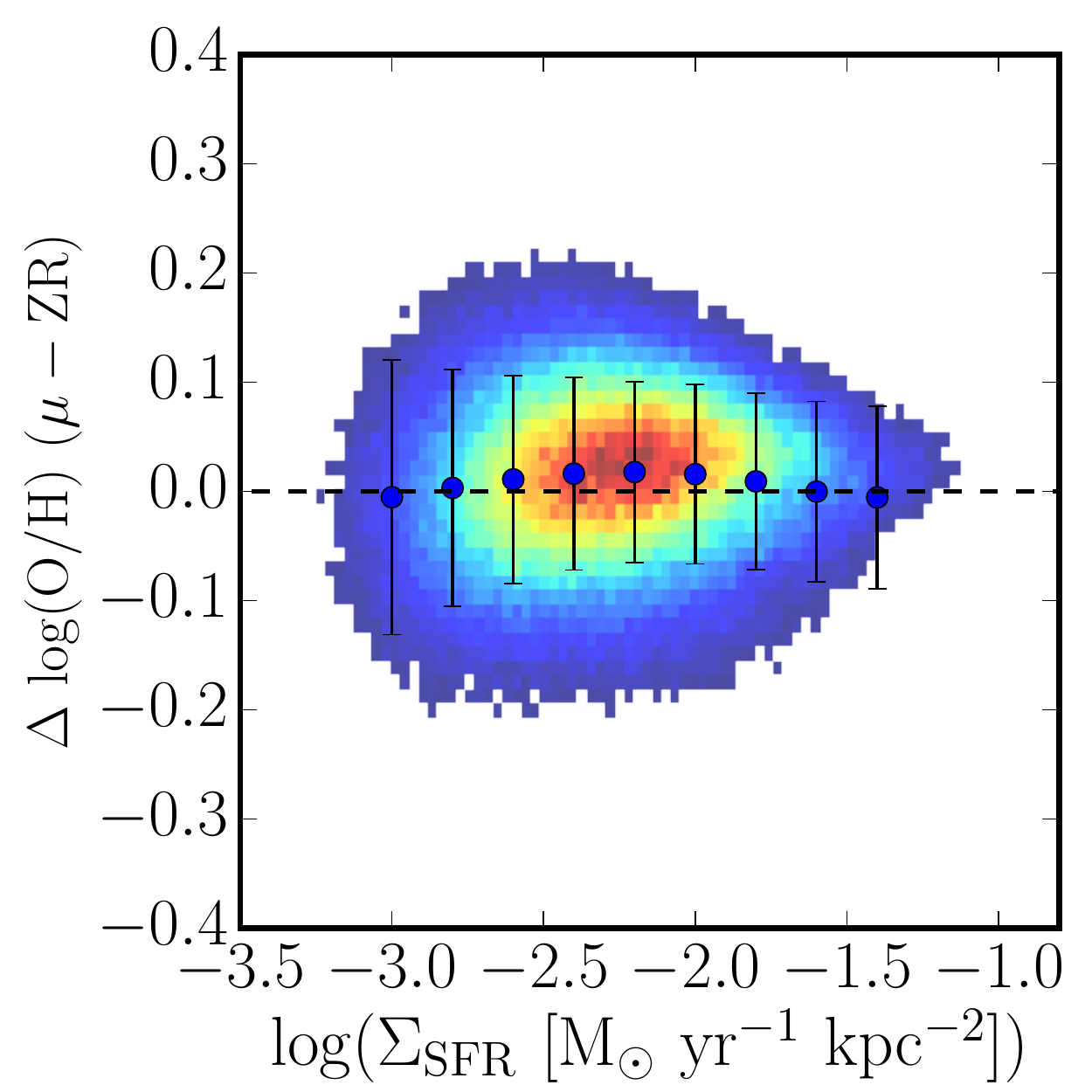}
\endminipage
\minipage{0.495\textwidth}
\includegraphics[width=\columnwidth]{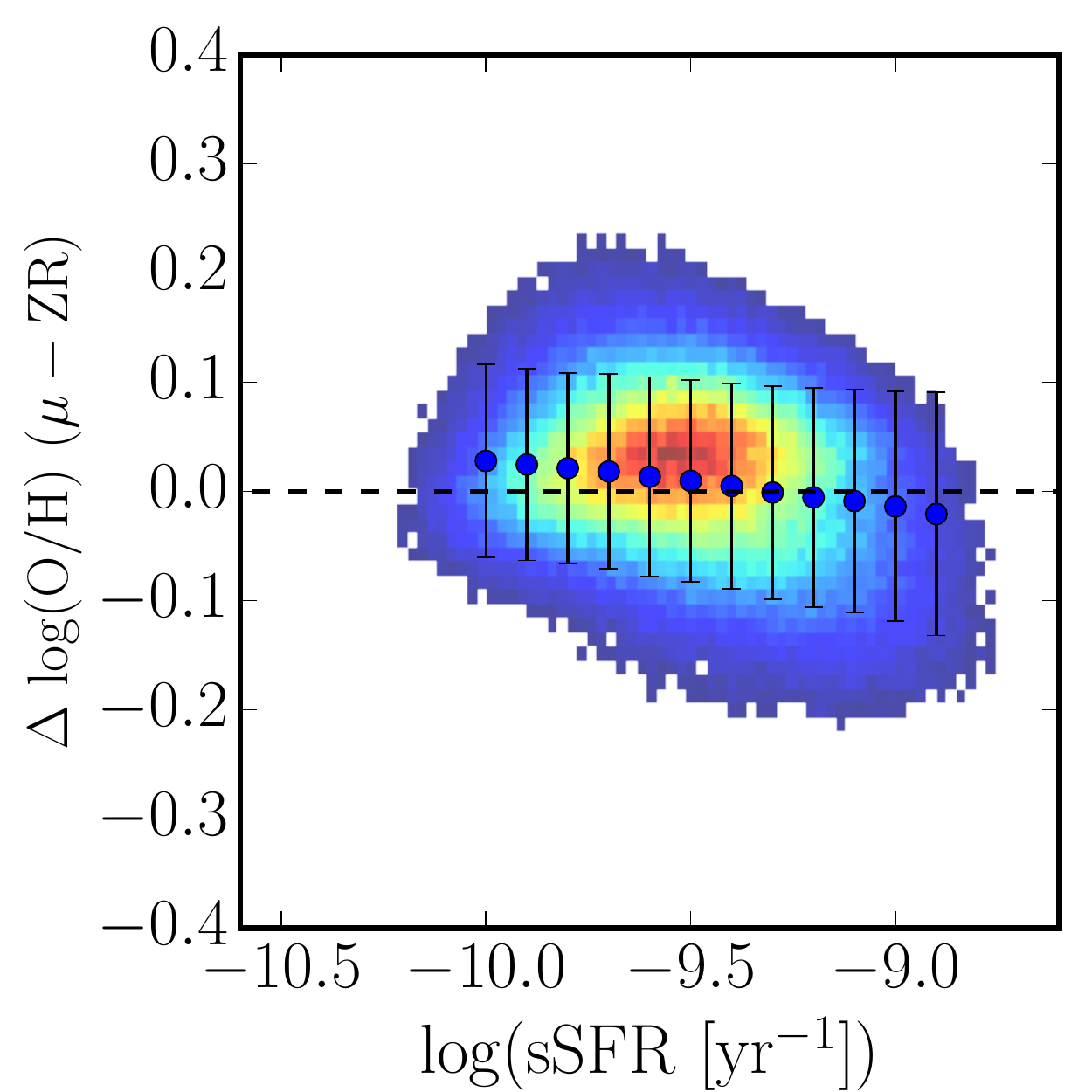}
\endminipage
\caption{Distributions of the residuals of the $\mu$-Z relation versus the SFR density ($\Sigma_{\rm SFR}$, left panel) and local specific SFR (sSFR = $\Sigma_{\rm SFR}/\Sigma_*$, right panel). The color distribution describes the density of spaxels with blue colors representing 15 spaxels per bin. The blue dots with error bars represents the median and standard deviation at bins of  $\Sigma_{\rm SFR}$ and sSFR for left and right panels, respectively. The dashed line shows the zero-line residuals.} 
\label{fig:muZres_SFR}
\end{figure*}
We have shown that there is a strong local inverse correlation between the total gas fraction $\mu$ and the Oxygen abundance (Z).  Since regions of high gas fraction will have higher star-formation rates, we might expect there to be a local correlation between Z and the specific star-formation rate (sSFR = $\Sigma_{\rm SFR}/\Sigma_*$) or $\Sigma_{\rm SFR}$. On the other hand, we have recently used MaNGA data to show that on global scales there is no significant correlation between SFR or sSFR and the residuals in the global M$_*$~-~Z relation \citep{BB_2017}.
 
To better understand the inter-relationships between $\mu$, SFR, and Z, in Fig.~\ref{fig:muZres_SFR}, we plot the residuals in the best-fit $\mu$-Z relation (Fig.\ref{fig:muZ}) as a function of both local sSFR and $\Sigma_{SFR}$. In the case of sSFR there is only a very weak trend (0.04 dex over a range of 1.2 dex in sSFR). There is no systematic trend in the case of $\Sigma_{SFR}$.  These results on local scales are consistent with the results on global scales presented in \cite{BB_2017}.

These results have a simple interpretation. The local ISM metallicity is most strongly dependent on the gas fraction (and secondarily on the local escape velocity).  Once these quantities are specified, the degree of local star-formation does not play a significant role in terms of predicting the metallicity. Any apparent correlation between Z and star-formation is a non-physical one induced by the mutual correlations of $\mu$ with both star-formation and Z.

\subsection{The Role of Local Stellar Surface Density}
 
In this paper we have examined the local dependence of the ISM metallicity (Z)  on the total gas-mass fraction $\mu$ and the escape velocity ($V_{\rm esc}$). We have focused on these two specific parameters because they can be most easily related to the parameters and processes invoked in simple models of the chemical evolution of galaxies. As we have shown above, at least one such model (a local version of the \cite{Lilly_2013} gas-regulator) does provide a statistically good, and physically reasonable, fit to our data.

In our previous paper \citep{BB_2016}, we found a very strong empirical local correlation between Z and the stellar surface density ($\Sigma_*$). Indeed, as we noted above, the scatter in this correlation is smaller than either the $\mu$-Z or $V_\mathrm{esc}$-Z relations presented in this paper. To close out this paper, we would like to briefly explore the connections between the $\Sigma_*$-Z relation and the results we have presented here.
 
It is well-established empirically that the star-formation history of galaxies (e.g., the luminosity-weighed mean age of the local stellar population) depends very strongly on $\Sigma_*$. For example, \citep{Kauffmann_2003} used the amplitude of the 4000 \AA\ break to show that the stellar age measured in the SDSS fiber  correlated more strongly with the global-average value of $\Sigma_*$ than with any other global property of the galaxy. Zheng et al. (2013) used multi-band imaging data to reach similar conclusions on a local scale\footnote{This does not hold in the outer disk beyond Petrosian R90 where the effects of  outward radial migration of stars becomes significant.}. More recently, \cite{Gonzalez-Delgado_2014} found a tight correlation between $\Sigma_*$ and the mean age of the stellar population at kpc scales in galaxies included in the CALIFA survey. 
 
High values of $\Sigma_*$ are characteristic of massive galaxies and the central regions of galaxies. Models of galaxy evolution  \citep[see,][]{Somerville_2015} find both that the progenitors of more massive galaxies form earlier and the centers of galaxies form first (e.g. the inside-out picture of galaxy formation). This then suggests a picture in which $\Sigma_*$ is a proxy for the redshift at which the dominant stellar population at that location formed or was assembled. In this picture, these early-forming dense regions were pre-destined to have low present-day gas fractions (the gas was mostly used up much earlier) and high escape velocities characteristic of the central regions of more massive galaxies. This suggests that the correlation between Z and $\Sigma_*$ is so good because $\Sigma_*$ encodes information about both $\mu$ and $V_\mathrm{esc}$.

\section{Conclusions}
\label{sec:Concl}
Thanks to the SDSS-IV MaNGA IFU survey, we have been able to study the spatially resolved properties for more than \mbox{9.2$\times$10$^{5}$} star forming spaxels in a sample of 1023 galaxies. Our goal was to use these data to test, on local scales, the general premises of models of chemical evolution in which the ISM metallicity is primarily set by the gas fraction (measuring the degree of chemical evolution) and the local escape velocity (which will regulate the rate at which gas and metals can be expelled by feedback from massive stars).

The main conclusions and results from this study are: 
\begin{itemize}
\item We presented the local relation between the gas fraction ($\mu$) and metallicity ($\mu$-Z relation) This tight relation ($\sigma \sim$ 0.09 dex) indicates that metallicity increases as gas fraction decreases.  

\item We constructed maps of the local escape velocity ($V_\mathrm{esc}$) based on modeling the galaxy rotation curve. With a larger scatter than the $\mu$-Z relation, we found that local metallicity also scales with $V_\mathrm{esc}$: spaxels with low escape velocity tend to be metal poor. 

\item We found weak but statistically significant systematic residuals in the $\mu$-Z relation as a function of $V_\mathrm{esc}$. We found stronger systematic residuals in the $V_\mathrm{esc}$-Z relation as a function of $\mu$. We concluded that gas fraction is the most important parameter in setting Z, but that the local escape velocity also contributes.

\item We fit local adaptations of the gas regulator and leaky box models of chemical evolution to the observed $\mu$-Z relation. The best fit leaky box model required unphysically large values for the mass-loading factor of the outflow. The best-fitting parameters from the gas regulator model suggest that local chemical composition is consistent with local outflows driven by the momentum supplied by massive stars and supernovae.

\item The scatter of the $\mu$-Z relation is similar to the one reported previously for the surface mass density - metallicity ($\Sigma_*$-Z) relation.  This latter quantity is a measure of the time of formation of the mass-dominant stellar at a given location, suggesting that the local metallicity, gas fraction and the escape velocity at the present epoch are pre-determined at much earlier times.

\end{itemize}

Our results indicate that both resolved gas fraction and ionized gas metallicity are the result of galaxy evolution occurring at local scales. This study also highlights the impact of local momentum-driven outflows into shaping the internal metallicity of star forming galaxies. 



\section*{Acknowledgements}
We thank the constructive comments from the referee. JKB thanks Molly Pepples and Lauren Corlies for their helpful discussion.  SFS thanks the ConaCyt programs IA-180125 and DGAPA IA100815
and IA101217 for their support to this project. Funding for the Sloan Digital Sky Survey IV has been provided by
the Alfred P. Sloan Foundation, the U.S. Department of Energy Office of
Science, and the Participating Institutions. SDSS-IV acknowledges
support and resources from the Center for High-Performance Computing at
the University of Utah. The SDSS web site is www.sdss.org.

SDSS-IV is managed by the Astrophysical Research Consortium for the 
Participating Institutions of the SDSS Collaboration including the 
Brazilian Participation Group, the Carnegie Institution for Science, 
Carnegie Mellon University, the Chilean Participation Group, the French Participation Group, Harvard-Smithsonian Center for Astrophysics, 
Instituto de Astrof\'isica de Canarias, The Johns Hopkins University, 
Kavli Institute for the Physics and Mathematics of the Universe (IPMU) / 
University of Tokyo, Lawrence Berkeley National Laboratory, 
Leibniz Institut f\"ur Astrophysik Potsdam (AIP),  
Max-Planck-Institut f\"ur Astronomie (MPIA Heidelberg), 
Max-Planck-Institut f\"ur Astrophysik (MPA Garching), 
Max-Planck-Institut f\"ur Extraterrestrische Physik (MPE), 
National Astronomical Observatories of China, New Mexico State University, 
New York University, University of Notre Dame, 
Observat\'ario Nacional / MCTI, The Ohio State University, 
Pennsylvania State University, Shanghai Astronomical Observatory, 
United Kingdom Participation Group,
Universidad Nacional Aut\'onoma de M\'exico, University of Arizona, 
University of Colorado Boulder, University of Oxford, University of Portsmouth, 
University of Utah, University of Virginia, University of Washington, University of Wisconsin, 
Vanderbilt University, and Yale University.
\bibliographystyle{mnras}
\bibliography{muZ_revised_TH}

\newpage

\appendix

\section{The Escape Velocity}
\label{app:EscapeVel}
\begin{figure*}
\includegraphics[width=\linewidth]{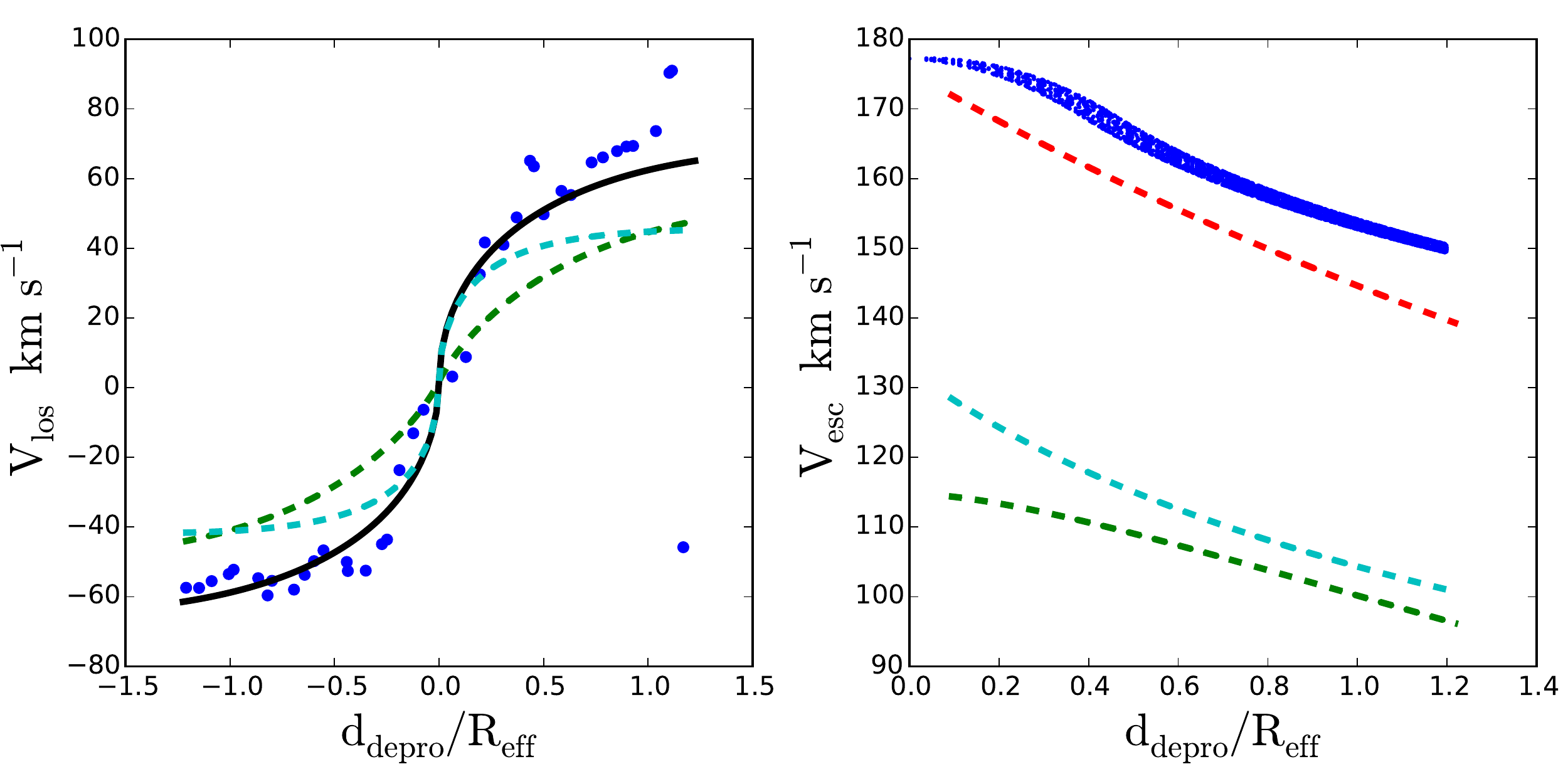}
\caption{Determination of the escape velocity using a two-component dynamical model for a disk-like galaxy included in our sample. Left panel: Blue points represent the rotation curve determined using the same procedure as described in Sec.~\ref{sec:HaVel}. The black solid line represents the best fit from the two-component dynamical model while green and cyan dotted lines represent the individual contributions to the rotational curve from the the axysimmetrical stellar disk and the spherical dark matter halo, respectively. Right panel: The blue points show the radial profile of the escape velocity as derived in Sec.~\ref{sec:Vesc} while the red dashed line represents the escape velocity from the best fit parameters using the two-component dynamical model. The green and cyan dashed lines represent the individual contributions to the escape velocity from the axisymmetric stellar disk and the spherical dark matter halo, respectively. In this example, the difference between the red line and blue points is $\sim$ 10 km s$^{-1}$.  } 
\label{fig:Example_kin_App}
\end{figure*}
In Sec.~\ref{sec:Vesc} we assume a spherically symmetric galactic potential for our sample of star-forming galaxies. In this Appendix we estimate the error introduced into the escape velocity from this assumption by using a more sophisticated dynamical model of a disk galaxy composed of an axisymmetric stellar disk within a spherical dark matter halo. For this particular example, we use a galaxy observed in the 127-fiber bundle setup to have a good spatial coverage, a Sersic index of $n < 2$ and axial ratio $0.25 < b/a < 0.75$ to avoid inclination effects. In the left panel of Fig.~\ref{fig:Example_kin_App} (see blue points) we plot the rotation curve derived using the same procedure as described in Sec.~\ref{sec:HaVel}

These rotation curves were then fit using a model rotation curve composed of an axisymmetric disk and spherical halo. The disk component was assumed to contain all of the baryonic matter and have an exponentially decreasing surface density:
\begin{equation}
\Sigma(R) = \frac{M_d}{2\pi R_d^2}e^{-R/R_d}
\end{equation}
where $M_d$ is the total mass of the disk and R is the radius in the plane of the disk (Binney $\&$ Tremaine, 2008). The disk mass in each case was taken from the NSA catalog. The NSA half life radius of each galaxy was used to obtain the disk scale radius by assuming a constant mass-to-light ratio and that the surface brightness and surface mass density each follow the same $e^{(R/R_d)}$ form. On the other hand, the halo component was modeled as a spherical NFW profile:
\begin{equation}
\rho(r) = \frac{\rho_0}{(r/R_h)(1 + r/R_h)^2}
\end{equation}
where r is the spherical radius (Navarro, Frenk, $\&$ White, 1996). The parameters $\rho_0$ and $R_h$ are taken as free parameters in our model. Both of these density profiles have analytically solvable potentials. The potential of the disk is given by (Binney $\&$ Tremaine, 2008): 
\begin{equation}
\Phi_{d}(R) = \frac{- G M_d y}{R_d}\bigg( I_0(y) K_1(y) - I_1(y)K_0(y)\bigg),  \; y = \frac{R}{2R_d} 
\end{equation}
where $I_n$ and $K_n$ are modified Bessel functions of the first and second kind, respectively. The potential of the halo is  (Binney $\&$ Tremaine, 2008):
\begin{equation}
\Phi_h(r, \rho_0, R_h) = -4\pi G \rho_0 R_h^2 \frac{\mathrm{ln}(1 + r/R_h)}{r/R_h}
\end{equation}
These potentials were then used to calculate the rotation curves $v_d(R)$ produced by the disk and $v_h(r, \rho_0, R_h)$ produced by the halo. We considered rotation curves and escape velocities within the plane of the disk where $r = R$. In addition to the disk and halo contributions, the model rotation curve also included a systemic velocity term to allow for errors in the systematic velocities used in deriving the rotation curve. Thus, the final model that was fit to our data is 
\begin{equation}
v(R,\rho_0,R_h, v_{sys}) = \sqrt{(v_d^2(R) + v_h^2(R,\rho_0,R_h))} + v_{sys}
\end{equation}
Once the halo and disk parameters were obtained from the best fit , the escape velocity to $R = \infty$  and halo mass $M_{100}$ (defined as the halo mass within 100 kpc) were calculated. In right panel of Fig.~\ref{fig:Example_kin_App} we show the comparison between the escape velocity derived using this two-component dynamical model (red-dashed line) and the one described in Sec.~\ref{sec:Vesc} (blue points). According to this comparison, the simplest model described in Sec.\ref{sec:Vesc} tends to overestimate the escape velocity in comparison to the two-component model. However, the difference of the order of $\sim$ 10 km s$^{-1}$ at different radii, which is a factor of $<$ 5\% of the estimated escape velocity from the single-component dynamical model in Sec.\ref{sec:Vesc}. This difference is similar in other disk-galaxies with similar characteristic as the one presented in this example.

%

\end{document}